\documentclass[prb,twocolumn,showpacs]{revtex4-1}

\usepackage{graphicx}
\usepackage{float}
\usepackage{color}
\usepackage{dcolumn}% Align table columns on decimal point
\usepackage{bm}% bold math
\usepackage{amssymb}
\usepackage{amsmath}
\usepackage{amsfonts}
\usepackage{bbm}
\usepackage{epsfig}
\usepackage{psfrag}
\usepackage{amsfonts}
\begin{document}

% the following line is for submission, including submission to the arXiv!!
%\hspace{5.2in} \mbox{Fermilab-Pub-04/xxx-E}

\title{Raman anomalies as signatures of pressure induced electronic topological and structural transitions in black phosphorus: Experiments and Theory}
\author{Satyendra Nath Gupta$^1$, Anjali Singh$^2$, Koushik Pal$^{2}$, Biswanath Chakraborti$^1$, D.V.S.Muthu$^1$, U V Waghmare$^2$  and A. K. Sood$^1$}
\thanks{To whom correspondence should be addressed}
\email{asood@physics.iisc.ernet.in}
\affiliation{$^1$Department of Physics, Indian Institute of Science, Bangalore-560012,India
$^2$Theoretical Sciences Unit, Jawaharlal Nehru Centre for Advanced Scientific Research, Bangalore-560064, India}      

\date{\today}

\begin{abstract}
We report high pressure Raman experiments of Black phosphorus (BP) up to 24 GPa. The line widths of first order Raman modes A$^1_g$, B$_{2g}$ and A$^2_g$ 
of the orthorhombic phase show a minimum at 1.1 GPa. Our first-principles density functional analysis reveals that this is associated with the anomalies 
in electron-phonon coupling at the semiconductor to topological insulator transition through inversion of valence and conduction bands marking a change 
from trivial to nontrivial  electronic topology. The frequencies of B$_{2g}$ and A$^2_g$ modes become  anomalous in the rhombohedral phase at 7.4 GPa, and 
new modes appearing in the rhombohedral phase  show anomalous softening with pressure. This is shown to originate from unusual structural evolution of black
phosphorous with pressure, based on first-principles theoretical analysis.

\end{abstract}
\pacs{78.30.-j, 71.30.+h, 71.15.Mb}

\maketitle
\section{Introduction}
The Black phosphorus (BP) is the most stable form among the many allotropic modifications of  phosphorus at room temperature and ambient pressure. 
It is an elemental  semiconductor with direct  band gap of $\sim$0.3 eV. Recently there is revival of interest in BP from the prospective of a layered 
material  which  allows us to study the exciting properties of a mono and  few layers of 
BP\cite{ling2015renaissance,chakraborty2016electron,li2014black,xia2014rediscovering,liu2014phosphorene,koenig2014electric,rodin2014strain,qiao2014high}. It
is found that the  band gap depends on the thickness of its film: 0.3eV for thickness $>4 nm$ and  1.2 eV for single layer thick 
BP \cite{zhang2014extraordinary}. This moderate band gap bridges the  gap between the semi-metallic graphene \cite{novoselov2005two,zhang2005experimental} 
and relatively large  band gap of transition metal dichalcogenide (1.5-2.5eV) \cite{mak2010atomically,splendiani2010emerging,jones2014spin} 
and makes it ideal for infrared optoelectronic application. Another interesting feature of BP is its anisotropic atomic structure. At ambient pressure, BP 
packs into orthorhombic lattice with a puckered  honeycomb structure as shown in Fig. \ref{structure}. Each phosphorus atom is bonded with two in-plane 
and one out of plane neighboring atoms. The puckered layers are bound together by weak van der Waals interactions. Due to this anisotropic atomic 
structure, the effective masses of carriers in BP along arm chair (ac) are lightest and that along the layer stacking (z) are lighter than that along the 
zigzag (zz)\cite{liu2015semiconducting}. Recent upsurge of interest in layered BP has arisen due to this unique  puckered 2D honeycomb lattice in 
each phosphorus layer, giving anisotropic electronic\cite{qiao2014high}, optical\cite{xia2014rediscovering,qiao2014high} and vibrational 
properties\cite{xia2014rediscovering}.

Further, the direct band gap ($\sim 0.3 eV$) of BP arises from the out-of-plane p$_z$-like orbitals of phosphorus and hence can be strongly 
modulated by pressure or c-axis strain. Not surprisingly, the effect of pressure was studied on crystal structure and electronic properties in 
the early work \cite{jamieson1963crystal,cartz1979effect,kikegawa1983x,burdett1982pressure,sugai1981pressure,vanderborgh1989raman,akahama1997raman}. 
It has anisotropic compressibility: lattice parameters a and c decrease more in comparison with the lattice parameter b with increasing pressure. At 
room temperature, the ambient pressure orthorhombic structure transforms to the rhombohedral structure at P$_{c2}$ $\sim$4.7 GPa and then to the simple 
cubic structure (sc) at P$_{c3}$ $\sim$ 11 GPa\cite{akahama1997raman}. These transition pressures are slightly higher if sample is cooled to liquid helium 
temperature first and then pressure is increased. The sc phase transforms to a simple hexagonal (sh) phase at higher pressure 137 GPa via an 
intermediate phase\cite{akahama1999simple}. Recently, Akahama et. al \cite{akahama2000structural} investigated the crystal structure of black phosphorus 
up to 280 GPa at room temperature using  x-ray-diffraction experiment and found that sh phase goes to the bcc structure at 262 GPa. The sc phase exhibits 
superconductivity at 4.7 K\cite{wittig1968superconducting,kawamura1984anomalous}. The superconducting temperature is slightly pressure dependent and 
increases with  increasing pressure. The superconducting state remains for several hours even if the pressure is removed. BP shows 
superconductivity with high transition temperature (10K) when it is cooled to 4K first and then pressure is applied. These experiments clearly 
establish  that high pressure can be used to tune  the material characteristic of BP significantly.

The electrical conductivity of black P is also  very sensitive to pressure \cite{Akahama 2001,Okajima 1984,akahama1986electrical}. Two anomalies 
have been reported at 1.7 and 4.2 GPa, the latter coinciding with  the  structural phase transition from orthorhombic to rhombohedral structure. The 
earlier work \cite{gong2015hydrostatic} reported that the band gap decreased linearly with a pressure coefficient of $\sim$ 0.17 eV/ GPa, closing 
at P$_{c1}$ $\sim$ 1.5 GPa. Recent magneto-resistance studies \cite{PhysRevLett.115.186403} have termed this transition as electronic topological 
transition (ETT) from semiconductor to 3D Dirac semi-metal due to band crossover near the Z-point in the Brillouin zone with linear dispersion. 
Interestingly, a colossal magneto-resistance of $\sim 80,000$ is seen at 2GPa at a field of 9T, similar to 2D Weyl semi-metals like WTe$_2$, making BP 
as an elemental semi-metal with increased electronic structure.

Pressure and strain have been used to change the relative strength of spin-orbit coupling and induce electronic topological transition (ETT) in 
materials like Sb$_2$Se$_3$ \cite{Bera}, b-As$_2$Te$_3$\cite{pal}, BiTeI\cite{xi}. Recently, Gong et al. \cite{gong2015hydrostatic} theoretically reported 
a pressure-induced Lifshitz transition  at a pressure of 1.2 GPa where black phosphorus undergoes a semiconductor to three-dimensional Dirac semimetal 
transition. By using first-principles density functional theoretical calculations within GGA and mBJ exchange-correlation functional, 
Gong et al. \cite{gong2015hydrostatic} showed that the band inversion occurs at the Z point at 1.2 GPa and robust three-dimensional Dirac points appear away from 
the Z point. Recently a theoretical study by Manjanath \textit{et. al}\cite{Manjanath} reported a reversible semiconductor to metal transition in 
bilayer phosphorene by applying normal compressive strain. In another work, Xiang \textit{et al. }\cite{PhysRevLett.115.186403, Akiba2015} reported 
colossal positive magneto-resistance in black phosphorus for P $\geq$ 1.2 GPa and observed a non-trivial $\pi$ Berry phase from the Shubnikov-de Haas 
oscillation measurement.~These observed changes in black phosphorus at pressure of 1.2 GPa were attributed to an electronic Lifshitz 
transition\cite{PhysRevLett.115.186403}.

The role of lattice in pressure-induced electronic topological transition has been evident in softening of an acoustic phonon mode at 
1.54 GPa in an earlier inelastic neutron scattering experiment\cite{PhysRevB.30.2410}, where anomalous compressibility and softening of 
longitudinal ultrasonic wave propagating along c-axis hint the relevance of  electron-phonon interaction to pressure induced transitions. High 
pressure Raman spectroscopy, a sensitive probe of lattice distortion under pressure, has been  used  in BP up to 13 GPa pressure, displaying 
changes in the pressure coefficients of Raman frequencies of a few modes as well as appearance of new modes at the structural transitions at 
P$_{c2}$ $\sim$ 4.7 GPa and P$_{c3}$ $\sim$ 10.5 GPa\cite{akahama1997raman}. Recently, Sasaki et al. \cite{Sasaki 2017} reported high 
pressure Raman study on few layers phosphorene and compared the results with bulk BP. It was found that monolayer and bilayer phosphorene 
show different behavior from the bulk sample. We revisit the high pressure Raman studies in this work with two objectives: (i) to see the 
Raman signature of electronic topological transition at P$_{c1}$  and (ii) to understand the phonon 
signatures using first-principles density functional theory (DFT). Our main results are (i) The linewidths of the first order 
Raman modes go through minimum at the ETT near P$_{c1}$ $\sim$ 1.1 GPa; (ii) Pressure dependence of phonon frequencies and 
appearance of new modes confirm the two structural transitions at  P$_{c2}$ $\sim$ 4.6 GPa and  P$_{c3}$ $\sim$ 11 GPa;  (iii) 
The first order modes B$_{2g}$ and A$^2_g$ show anomalous softening in the rhombohedral phase; (iv) All the three 
Raman modes of the orthorhombic structure vanish above $\sim$ 11 GPa in the sc phase; (v) The new modes 
appearing in the rhombohedral phase show anomalous decrease in frequency with increasing pressure 
up to 24 GPa; (vi) Using first-principles analysis we show that the transition at the lowest pressure is an 
electronic topological transition  to a phase with a strong $Z_2$ topology, while  the anomalies at higher pressure 
have a structural origin. 
(vii) The new modes labeled N1, N2 and N3 observed in the rhombohedral phase have been quantitatively analyzed 
theoretically for the first time, capturing their anomalous softening.  The modes N1, N2 and N3 have been assigned to $E_g^1$, $A_{1g}$ and $E_g^2$ modes respectively. The modes N1 and N2 in the sc phase are assigned as X and M point acoustic modes, respectively.

\section{Experimental Details}
Raman measurements were carried out at room temperature in back scattering configuration with micro-optical system equipped 
with Horiba 800 spectrometer, Peltier-cooled charge controlled 
detector and 532 nm diode laser. Pressure was generated using a diamond anvil cell (DAC) using 4:1 methanol: 
ethanol pressure transmitting medium. A thin platelet of dimension $\sim 100 \mu m$ cleaved from single crystals of BP, 
(from M/S Smart Elements) was placed into a stainless steel gasket with sample chamber hole of $\sim 200\mu m$ inserted 
between the diamonds along with a ruby chip for pressure calibration. Laser power ($<5 mW$) was kept low enough 
to avoid damage to the sample.

\section{Computational Details}
Our first-principles calculations are based on density functional theory (DFT) as implemented in the Quantum ESPRESSO package\cite{qe}, 
with the interaction between ionic core and valence electrons  modeled with norm-conserving pseudopotentials \cite{SG, CH}. 
The exchange-correlation energy of electrons is treated within a generalized gradient approximation (GGA) with a functional 
form parameterized by Perdew, Burke and Ernzerhof \cite{PBE}. We use an energy cutoff of 55 Ry to truncate the plane wave 
basis used in representing Kohn-Sham wave functions, and energy cutoff of 220 Ry for the basis set to represent charge density. 
Structures are relaxed to minimize energy till the magnitude of Hellman-Feynman force on each 
atom is less than 0.03 eV/\AA. We include van der Waals (vdW) interaction with the parametrization given in Grimme scheme \cite{Grimme}. In 
self-consistent Kohn-Sham (KS) calculations of configurations of bulk black phosphorous with orthorhombic and rhombohedral unit cell, the 
Brillouin zone (BZ) integrations were sampled with a uniform mesh of 18x16x16 and 16x16x16 k-points respectively, and 32x32x32 mesh 
of k-points was used in calculation of electron-phonon coupling of both the structural forms. Phonon spectra and dynamical matrices 
at $\Gamma$-point (q= (0, 0, 0)) as a function of lattice constant (or pressure) were determined using density functional linear 
response as implemented in Quantum ESPRESSO(QE)\cite{qe}, which employs the Green's function method to avoid explicit calculations of 
unoccupied Khon-Sham states. Since DFT typically underestimates the electronic bandgap, we have used HSE functional\cite{HSE} 
to estimate the gaps accurately, with the mixing parameter of 0.25 and reciprocal space integration sampled on a 4x4x4 mesh of k-points.
We have fully relaxed the orthorhombic structure of 
BP with HSE calculations maintaining the experimental c/a ratio. We find that the error in resulting cell volume (relative to experiments) 
is negligibly zero (Fig.~\ref{gap} (a)). Fitting a quadratic function to energy vs. volume curve, we used -$\frac{dE}{dV}$ to estimate 
pressures within HSE calculations. We determine electronic bandgap at Z-point as a function of volume, and used the point of minimum 
bandgap to  estimate the transition pressure within HSE calculations (Fig.~\ref{gap} (b)).

Z$_2$ topological invariants can be defined using the notion of time reversal polarization \cite{soluyanov, joydeep} derived in terms of hybrid Wannier 
charge centers (WCCs) \cite{fu}. In a time-reversal invariant system, electronic bands always come in time-reversed pairs 
(let's say I \& II denote such pairs). Then the Z$_2$ invariant in  a time-reversal invariant plane is given by \cite{soluyanov},

\begin{equation}
(\sum_{n}[\bar{x}_n^{I}(T/2) - \bar{x}_n^{II}(T/2)]-\sum_{n}[\bar{x}_n^{I}(0)- \bar{x}_n^{II}(0)])~ mod~ 2,
\end{equation}

where $\bar{x}_n = 
\frac{i}{2\pi}\int_{-\pi}^{\pi}dk<u_{nk}|\frac{\partial}{\partial_k}|u_{nk}>$ is the Wannier charge center calculated at 
t=0 and t=T/2 planes which are invariant under time reversal symmetry, where T represents the period of a full cyclic adiabatic evolution. In the 
Brillouin zone of a periodic crystal T is equivalent to a reciprocal lattice vector which defines the periodicity in the reciprocal space. The 
topological invariant of a plane is non-zero if the WCCs switch pairs under an adiabatic evolution in the half-cycle, which can be easily tracked 
by seeing evolution of the mid-point of the largest gap (as marked by blur rhombus in Fig.~\ref{wcc}) between  two adjacent WCCs   with t $\in$ [0, T/2] 
in the half-cycle \cite{soluyanov}. In this case, the rhombuses  exhibits abrupt jumps in their half-cyclic evolution and cross odd number 
of WCCs \cite{soluyanov}. We have calculated the strong topological index ($\nu_0$) by taking the sum (modulo 2) of the topological invariants 
calculated at k$_z$=0 and k$_z$=0.5 planes in the Brillouin zone of black phosphorous using Z2Pack code \cite{gresch}. At each of these planes, WCC, 
calculated along k$_x$ direction,  evolve along the k$_y$ direction (parameterized with time, t).

\section{Results and Discussion}
\subsection{Experimental Results}

\begin{figure*}
\includegraphics[width=0.9\textwidth]{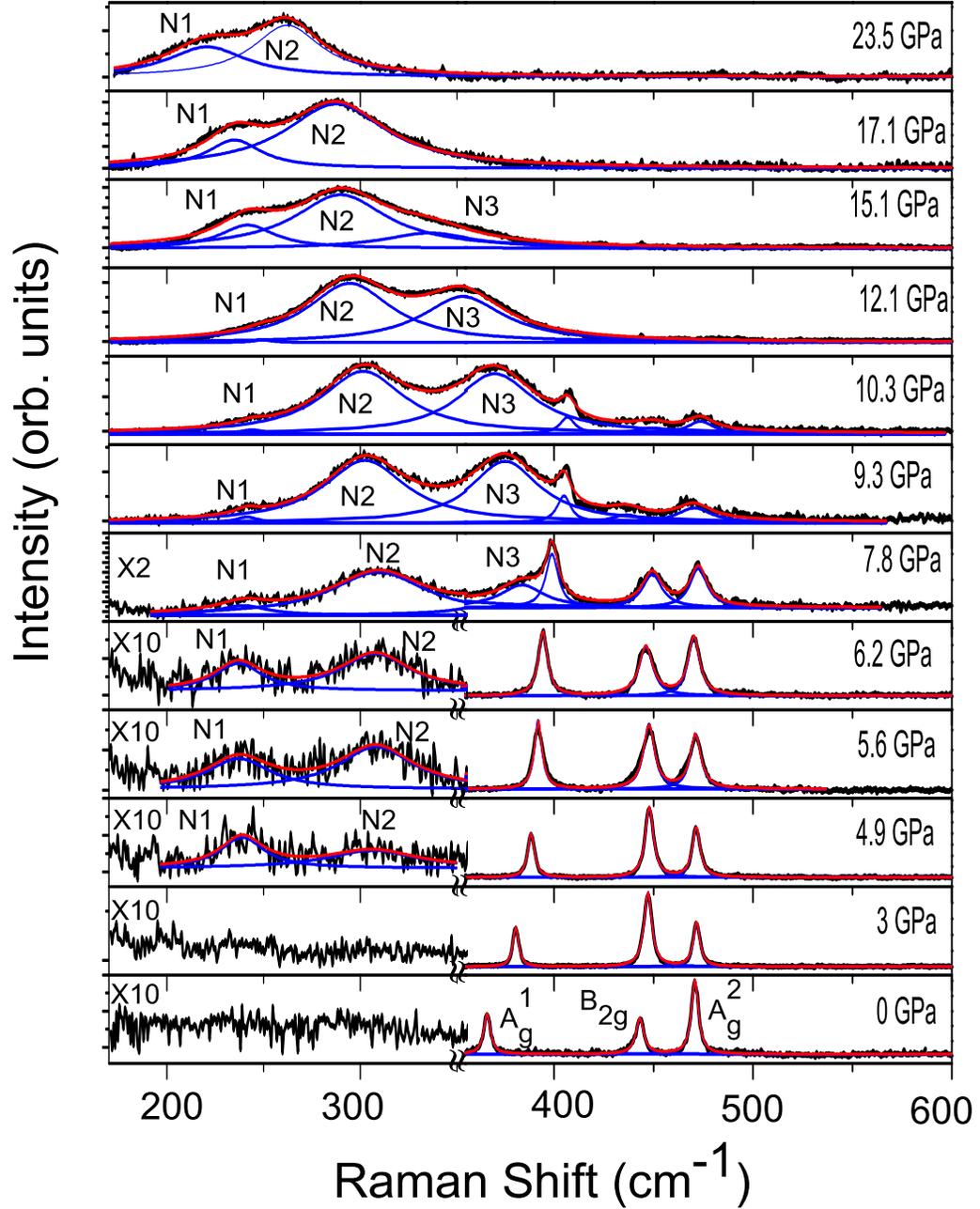}
\caption{Pressure evolution of Raman spectra. The solid lines (red and blue) are the Lorentzian fits to the experimental 
data (black). N1, N2 and N3 are the new modes.
\label{Raman}}
\end{figure*}

\begin{figure}
 \includegraphics[width=0.48\textwidth]{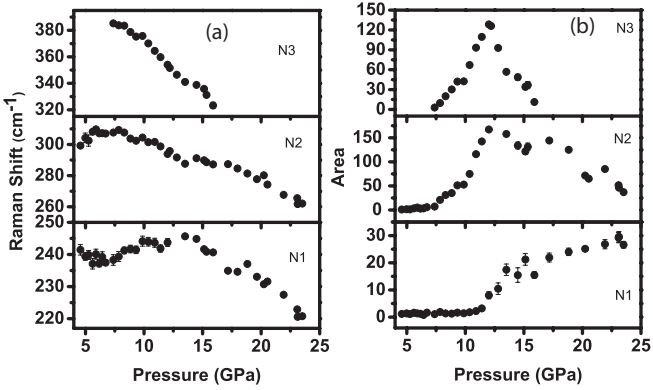}           
 \caption{Pressure dependence of phonon frequencies and area of N1, N2 and N3 modes.}
\label{N123}
\end{figure}

The unit cell of BP contains 4 atoms which gives 12 normal modes. According to factor group analysis, at ambient pressure the Brillouin zone center 
point phonon modes are 2A$_g$+B$_{1g}$+B$_{2g}$+2B$_{3g}$+A$_u$+2B$_{1u}$+2B$_{2u}$+B$_{3u}$: out of which  2A$_g$, B$_{1g}$, B$_{2g}$ and 2B$_{3g}$ 
are Raman active. Experimentally three Raman modes  A$^1_g$, B$_{2g}$ and A$^2_g$ were observed at 360, 438 and 460 cm$^{-1}$ respectively as shown in 
lower panel of Fig.~\ref{Raman}. The other two previously reported\cite{akahama1997raman} Raman modes  B$_{1g}$ (192 cm$^{-1}$) and  B$_{3g}^1$ 
(227 cm$^{-1}$) are very weak and difficult to separate  from the background. The phosphorus atoms vibrate along the out of plane (z) for 
the A$^1_g$ (360 cm$^{-1}$) mode, along armchair (y) for the B$_{2g}$ (438 cm$^{-1}$) mode and zig zag (x) directions for the A$^2_g$ (460 cm$^{-1}$) mode. 
Fig.~\ref{Raman} shows the pressure evolution of the Raman spectra of BP at a few representative pressures. The Lorentzian line shapes are used to 
fit the  Raman spectra at different pressures to obtain phonon frequencies and full width at half maximum (FWHM). From Fig.~\ref{Raman}, it can be seen
that the intensities of A$^1_g$, B$_{2g}$ and A$^2_g$ Raman modes of the orthorhombic structure decrease gradually with pressure and disappear 
completely at 11 GPa. Further, two new Raman modes labelled as N1 and N2 at 240 and 300 cm$^{-1}$ respectively, emerged at 4.6 GPa and remain till 
24 GPa. Another new mode labelled as N3 at 370 cm$^{-1}$ arises at 7.4 GPa and diminishes at 15.9 GPa. To shed more light on the evolution of modes 
N1, N2 and N3 with pressure, we examine the frequencies and area of N1, N2 and N3 peaks as a function of pressure (see Fig.~\ref{N123}). It is clear 
that the intensity of N1 mode increases slowly with increasing pressure till 10 GPa, and rapidly beyond 10 GPa. The intensity of N2 and N3 modes 
increases with pressure till 13 GPa, and decreases beyond that. The phonon frequencies of N1, N2 and N3 modes  soften with pressure.

\begin{figure*}
\includegraphics[width=0.9\textwidth]{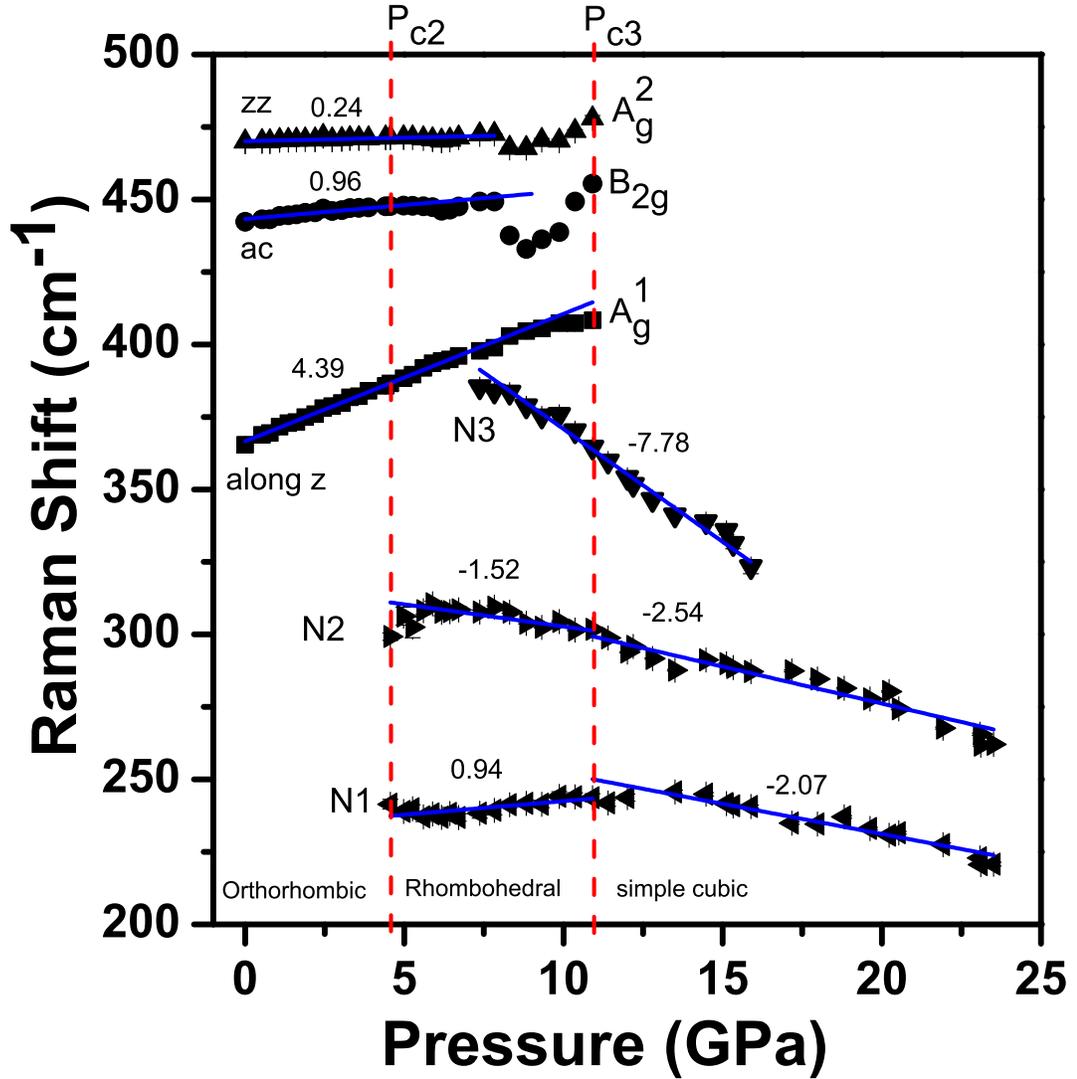}
\caption{Pressure dependence of  phonon frequencies. The 
vertical dashed lines indicate the two structural phase transition pressures. The solid blue lines are the linear fit. The 
slope d$\omega$/dP in units of cm$^{-1}$ /GPa is given near the lines.
\label{Figure-Pressure dependence of phonon frequencies}}
\end{figure*}

\begin{figure*}
\includegraphics[width=0.9\textwidth]{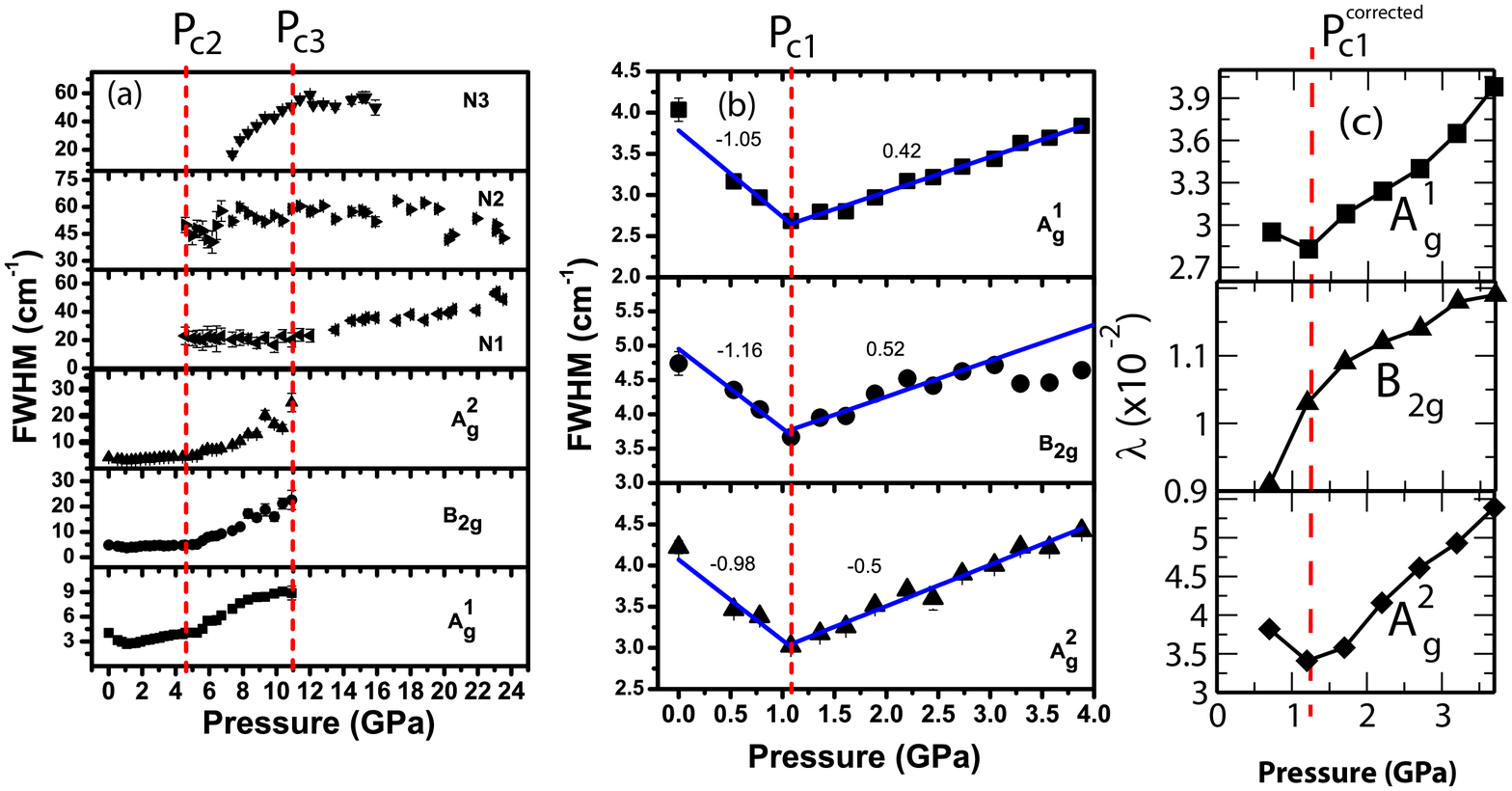}
\caption{(a) Pressure evolution of FWHM of  Raman modes. The vertical dashed 
lines indicate the two structural phase transition pressures. (b) Magnified plot FWHM 
in the pressure range 0 to 4 GPa. The vertical dashed line indicates the semiconductor to topological insulator 
phase transition pressures. The solid blue lines are linear fit. The slope d$\omega$/dP in units of cm$^{-1}$ /GPa is given near 
the lines. (c) Variation in the calculated electron-phonon coupling of Raman active modes $A^{1}_{g}$, $B_{2g}$ and $A^{2}_{g}$ of BP with 
pressure in orthorhombic phase calculated using GGA functional. The pressure axis has been corrected as explained in the text. The vertical dashed 
line indicates the semiconductor to topological insulator phase transition pressures.
\label{Figure-Pressure evolution of FWHM}}
\end{figure*}

Fig.~\ref{Figure-Pressure dependence of phonon frequencies} shows the pressure dependence of  phonon frequencies. The solid blue lines are the linear 
fit to the experimental data points.  The vertical dashed red lines indicate two structural phase transitions from orthorhombic to rhombohedral and 
from rhombohedral to simple cubic phase. Fig.~\ref{Figure-Pressure dependence of phonon frequencies} shows that the frequencies of A$^1_g$, B$_{2g}$ 
and A$^2_g$ modes show increase with pressure in orthorhombic phase. The slope for the A$^1_g$ (S=d$\omega$/dP) is more as compared to that for 
the B$_{2g}$ and A$^2_g$ modes. Noting that the eigen vectors of the A$^1_g$ mode are perpendicular to the puckered layers, larger pressure 
derivative of frequency (S)  signifies that van der Waals bond can be greatly compressed. Fig.~\ref{Figure-Pressure dependence of phonon frequencies} 
also shows that all the three modes of the orthorhombic phase (A$^1_g$, B$_{2g}$ and A$^2_g$) persist even beyond 4.6 GPa. This can be due to 
two possibilities (i) that the orthorhombic to rhombohedral transformation is completed only at $\sim$ 11 GPa; (ii) the Raman active modes of 
the rhombohedral phase are at the frequencies close to the values in the orthorhombic phase. The latter possibility is ruled out as the frequencies of Raman modes in  rhombohedral phase (to be discussed latter)  are not close to the above three modes. In rhombohedral phase 
at 7.4 GPa, A$^2_g$ and B$_{2g}$ modes  soften anomalously (the effect is more for B$_{2g}$ mode). Further 
Fig.~\ref{Figure-Pressure dependence of phonon frequencies} shows that the Raman modes N1, N2 and N3 show anomalous softening with pressure in 
rhombohedral as well as in simple cubic phase. It is clear from Fig.~\ref{Figure-Pressure dependence of phonon frequencies} that the slope S for the 
N3 mode is more than N1 and N2 modes.  Fig.~\ref{Figure-Pressure evolution of FWHM}(a) shows the effect of pressure on FWHM of 
the Raman modes. The  range from 0 to 4 GPa  has been magnified  in  Fig.~\ref{Figure-Pressure evolution of FWHM}(b). The solid blue 
lines are the linear fit. It is clear from Fig.~\ref{Figure-Pressure evolution of FWHM}(b) that the FWHM of all the three Raman modes 
show a minimum at 1.1 GPa which coincide with the ETT transition pressure. Another thing to note is that the line width of N1, N2 and N3 modes 
are very large. We don't fully understand the origin of the large linewidths of these modes in the rhombohedral phase. One possible origin 
could be the high disorder in the rhombohedral phase.  We now present our theoretical calculations to understand the anomalies in phonon 
spectra with pressure observed in the experiments. 

\begin{figure}
\includegraphics[width=0.5\textwidth]{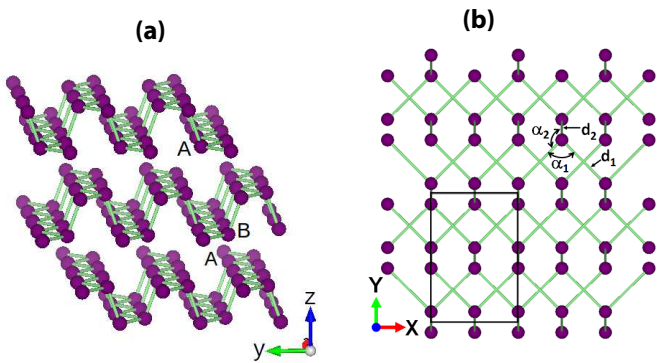}
\caption{Side (a) and top (b) views of the layered structure of bulk black phosphorous. $d_1$ and $d_2$ are the 
P-P bonds lengths. $d_1$ is the distance between two P atoms in a plane while $d_2$ is the nearest neighbor distance. $\alpha_1$ is 
the bond angle between two $d_1$'s and the bond angle between $d_1$ and $d_2$ is $\alpha_2$.
  \label{structure}}
 \end{figure}

\begin{figure*}
\includegraphics[width=0.6\textwidth]{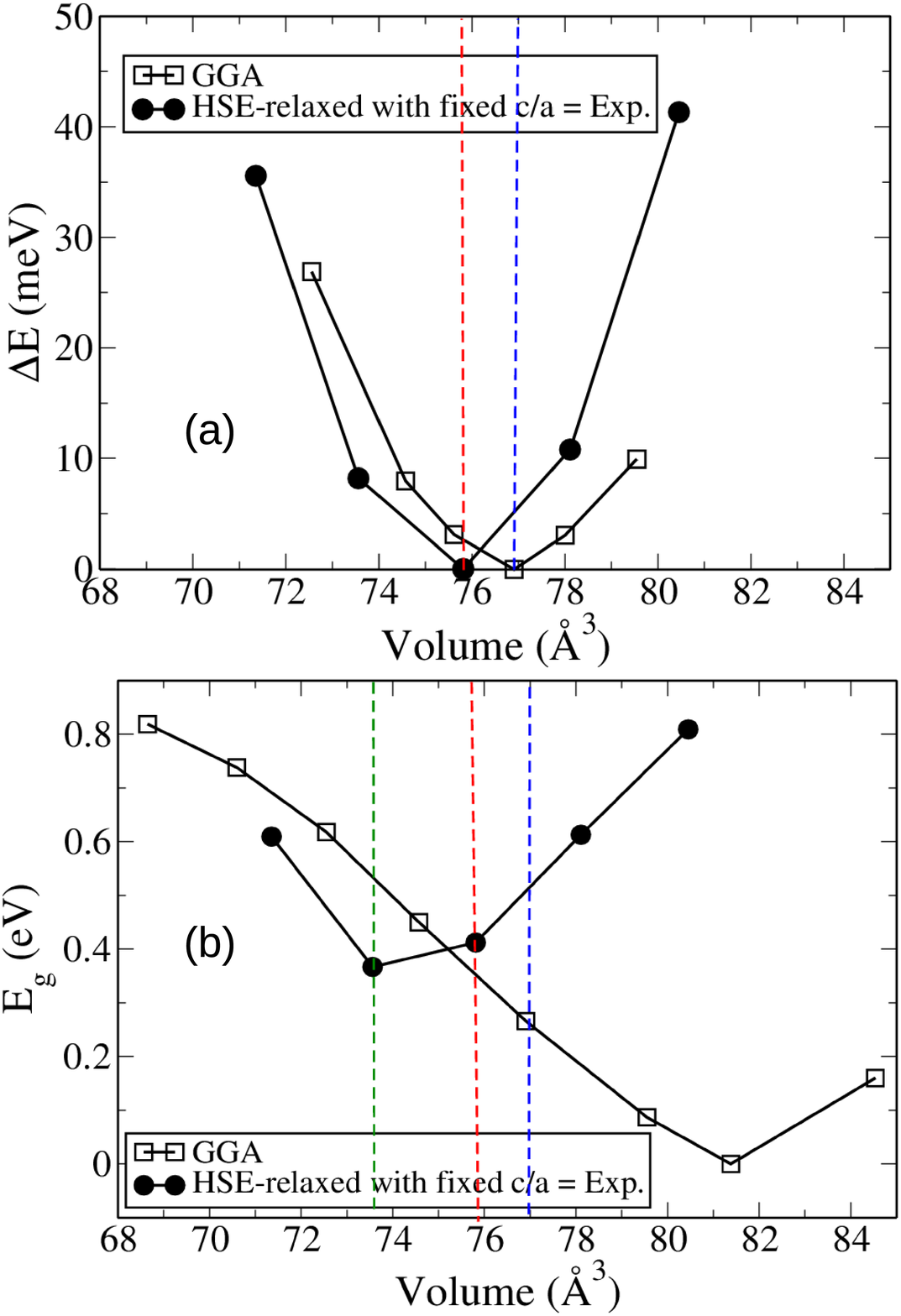}
\caption{Change in energy (a) and bandgap at Z-point of Brillouin zone (b) with volume obtained from 
calculations with GGA and HSE functionals. 
Red and blue dashed lines mark the equilibrium structure of BP in orthorhombic phase. Note that green dashed 
line highlights the volume at which bandgap of HSE-relaxed structure exhibits a minimum, making the electronic topological transition.}
\label{gap}
\end{figure*}

 \subsection{Theoretical calculations}
\subsubsection{ Low Pressure Transition}
\begin{figure*}
\includegraphics[width=0.9\textwidth]{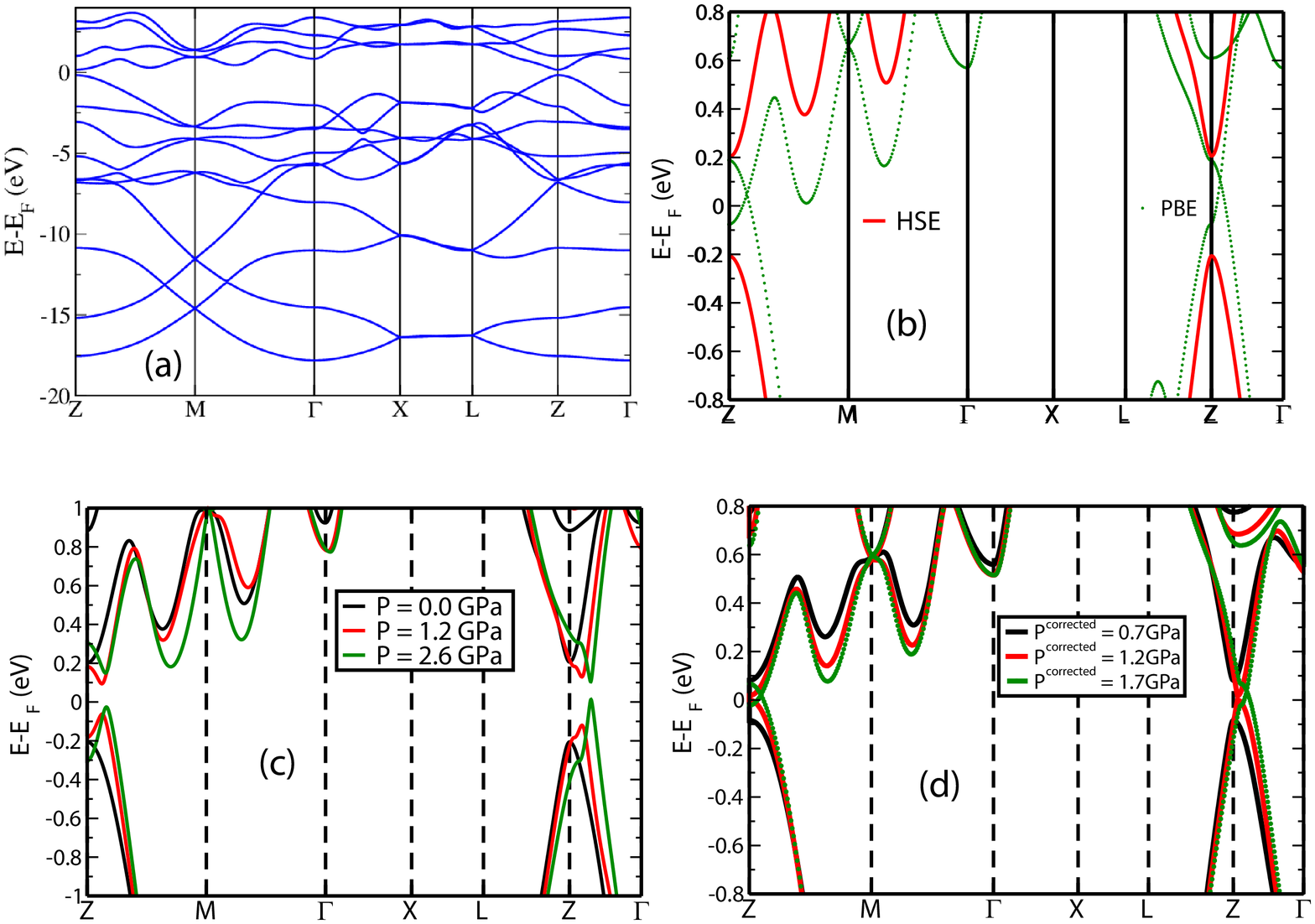}
\caption{Electronic structure of BP in orthorhombic phase. (a) Electronic structure of BP at $P= 0$ GPa calculated with HSE functionals. 
Note that we used Wannier functions\cite{Wannier} to plot band structure with HSE functionals. (b) Comparison of electronic structure near 
the gap obtained with HSE and PBE functionals at $P= 0$ GPa. BP exhibits a gap of 0.33 eV at Z-point, which is captured 
correctly with HSE functionals whereas PBE calculations gives no gap. (c) Electronic structure near the gap obtained with HSE functionals 
at different pressures. (d) Electronic structure of BP calculated with PBE-GGA functional at different pressures. The pressures have been corrected 
as explained in the text. At P$^{corrected}$ = 0.7 GPa, calculations with PBE functionals show a gap which closes at P$^{corrected}$=1.2 GPa. With further 
increase in pressure, bandgap closes in vicinity of Z-point at P$^{corrected}$ = 1.7 GPa.}
\label{bandstructure}
\end{figure*}

\begin{figure}   
\includegraphics[width=0.5\textwidth]{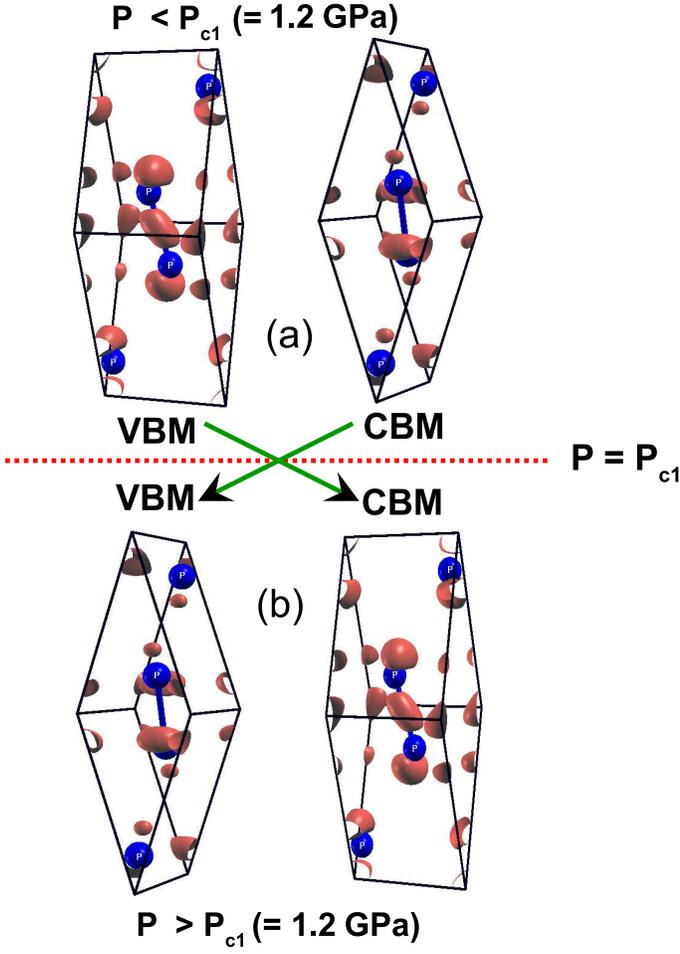}
\caption{Band inversion during pressure induced electronic topological transition (ETT) in BP. Isosurfaces of charge densities 
associated with electronic states at valance band maximum (VBM) and conduction band minimum (CBM) at $\Gamma$-point (a) before 
and (b) after the ETT revealing band inversion across this transition.
\label{bandinversion}}
\end{figure}

BP is a semiconductor with a gap of 0.40 eV at Z-point (0.0, 0.0, 0.5) as estimated with HSE functional 
(Fig.~\ref{bandstructure}(a)), which is in good agreement with the experiment ($E_g$ = 0.35 eV). We note 
that there is no band gap in the structure at 0 GPa calculated with GGA functional based theory (Fig.~\ref{bandstructure}(b)).
Raman experiments ( Fig.~\ref{Figure-Pressure evolution of FWHM}(b)) showed non-monotonic dependence of FWHM of 
$A^1_g$, $B_{2g}$ and $A^2_g$ modes on pressure, displaying a minimum at 1.1 GPa.  
Since the anomalous change in FWHM can be an indicator of electronic topological transition (ETT) \cite{Bera}, 
and given the experimental observations of Xiang \textit{et al.} \cite{PhysRevLett.115.186403} and theoretical 
study of Gong et al.\cite{gong2015hydrostatic}, we first studied electronic structure of black phosphorus as a function 
of hydrostatic pressure using highly accurate HSE calculations.  
Our calculations show that the band inversion occurs at Z point at P$_{c1}$ $\sim$ 1.2 GPa (Fig.~\ref{bandinversion}) following which 
the band gap opens up above P$_{c1}$ $\geq 1.2$ GPa (Fig.~\ref{bandstructure}(c)), which is in contrast to the earlier DFT study 
(within GGA and mBJ functional) where persistence of robust band crossing is reported
after band inversion \cite{gong2015hydrostatic}. In an electronic topological transition the bandgap decreases with 
increasing pressure, closes at $P_c$ and then increases (opens up) with further increase in pressure \cite{Bera}. 
We have plotted the bandgap of BP at Z-point (the point at which band inversion occurs) in the vicinity of critical pressure, 
which clearly shows bandgap reduces and then increases as a function of volume (Fig~\ref{gap}(b)) \textit{i.e.}, the bandgap exhibits 
a minimum value at $P_{c1}$ and then increases. This is associated with closure of bandgap at a k-point near Z-point. 
To probe this, we have used Wannier functions to interpolate and plot 
HSE band structure. Oscillating behavior in the bandgap along Z-M and Z-$\Gamma$ is due to errors in interpolation of bands with 
Wannier functions near the gap closure point, where band energies are non-analytic functions of k. We find that the HSE  
bandgap minimum is at volume lower than of the equilibrium structure. Hence ETT occurs at positive pressure ($P_{c1}$ $\sim$ 1.2 GPa), 
which is comparable with the pressure of anomalies seen in experiments. Based on the HSE calculations, our estimate of the 
transition pressure  is $P_{c1}$ $\sim$ 1.2 GPa, 
in reasonable agreement with theoretically predicted transition pressure (0.6 GPa) by 
Ruixiang \textit{et. al} \cite{fei}, and also close to the pressure of observed Raman anomalies (P$_{c1}$ $\sim$1.1 GPa).

\begin{figure}  
\includegraphics[width=0.5\textwidth]{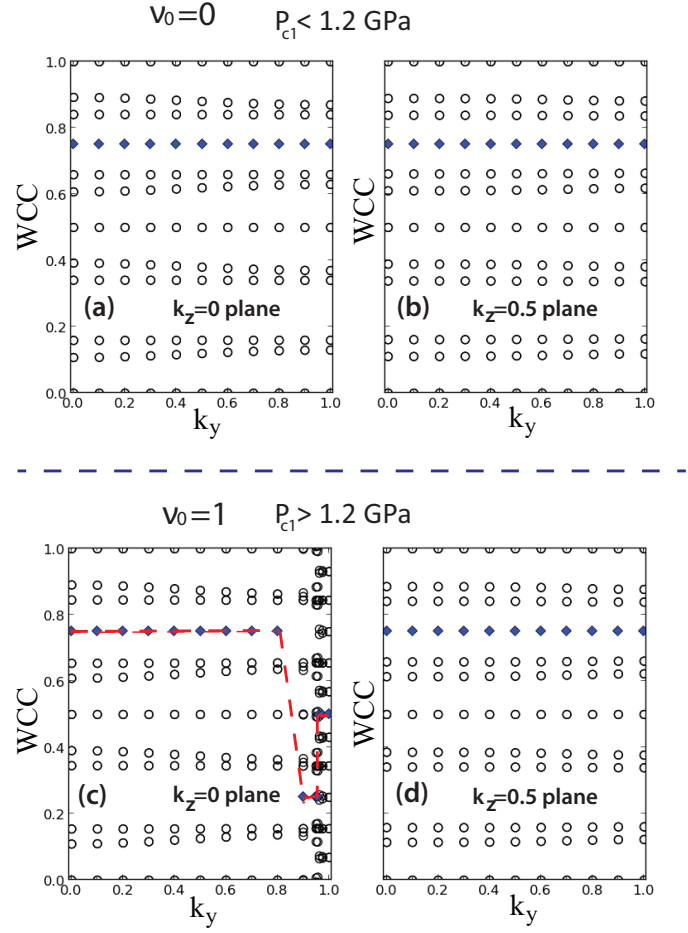}
\caption{Calculation of Z$_2$ topological index ($\nu_0$) before (P $<$1.2 GPa) and after (P $>$1.2 GPa) the phase transition. 
Evolution of hybrid Wannier charge centers (WCCs) along k$_y$-direction (marked by circle) and their largest gap function (blue rhombus) 
for (a) $k_z$ =0 plane and (b) $k_z$=0.5 plane. It is clear that both $k_z$ = 0 and $k_z$=0.5 planes are topologically trivial 
as Z$_2$ invariants for both the planes are zero, hence strong topological index $\nu_0$=0. At P $>$1.2 GPa, Z$_2$ invariant 
for $k_z$ = 0 plane is 1, as evident from the discontinuous jump (indicated by red dotted line) of the gap function along  k$_y$ (c), 
whereas Z$_2$ invariant is zero for $k_z$=0.5 plane (d), hence the strong topological index $\nu_0$ at this pressure is 1.
\label{wcc}}
 \end{figure}

Although earlier studies \cite{gong2015hydrostatic, PhysRevLett.115.186403} reported pressure induced electronic topological transition 
(Lifshitz transition) in black phosphorus following band inversion near P = 1.2 GPa, the nature of electronic topology has not 
been clearly elucidated. To investigate the electronic topology of black phosphorus, we determined the $Z_2$ topological 
invariant ($\nu_0$) as a function of pressure (Fig.~\ref{wcc}). 
The evolution of hybrid Wannier charge centers (WCCs) along k$_y$-direction is marked by circle while their largest gap 
function is marked by blue rhombus for  $k_z$ =0 plane and  $k_z$=0.5 plane. It is clear from Fig.~\ref{wcc} (a),(b) that for P $<$ 1.2 
GPa both $k_z$ = 0 and $k_z$=0.5 planes are topologically trivial as Z$_2$ invariants for both the planes are zero, and hence 
strong topological index $\nu_0$=0. However, at P $>$1.2 GPa (Fig.~\ref{wcc} (c),(d)), Z$_2$ invariant for  $k_z$ = 0 plane is 1, 
as evident from the discontinuous jump of the gap function along  k$_y$, whereas Z$_2$ invariant is zero for $k_z$=0.5 plane, 
and hence the strong topological index $\nu_0$ of the state at this pressure is 1. Thus we find that $\nu_0 = 0$ for $P< 1.2$ GPa, 
whereas $\nu_0 = 1$ for $P \geq 1.2$ GPa signifying $Z_2$ topological insulating nature of black phosphorus above 1.2 GPa. Thus, 
the minimum in FWHM of Raman active modes at (P$_{c1}$) is correlated with the pressure induced electronic topological phase transition. 
The observation of non-trivial $\pi$ Berry phase and colossal magneto-resistance for P $>$ 1.2 GPa in experiment \cite{PhysRevLett.115.186403} 
are also consistent with our finding that the BP becomes topologically non-trivial for P $>$ 1.2 GPa. The discrepancy between 
our calculated transition pressure (1.2 GPa) and previously reported ones \cite{fei} (0.6 GPa) falls within the typical 
errors of DFT calculations.

Calculations of phonon frequencies and electron-phonon coupling are not straight-forward with the HSE functionals 
because (i) DFT-LR (DFT linear response) implementation with HSE functional is not available presently and (ii) frozen phonon
method with fine mesh of k-points make HSE calculation very computationally intensive. Thus we have used GGA functional in understanding 
the observed phonon anomalies. We correct the pressures estimated in GGA calculations with a shift of 2.7 GPa, to match the transition
pressure estimated with HSE calculations in the orthorhombic phase.  
Our optimized structural parameters of orthorhombic structure, a, b, c, $d_1$, $d_2$, $\alpha_1$ and $\alpha_2$ are  4.42{\AA},
3.32{\AA}, 10.46{\AA}, 2.224 {\AA}, 2.244 {\AA}, 96.34$^\circ$ and 102.09$^\circ$ respectively, in good agreement
with the experimental \cite{Akai 1989} and earlier reported theoretical\cite{Svane 2012} values.
Here $d_1$ is the bond length of in-plane P atoms,  $d_2$ defines the distance out-of-plane P atoms; $\alpha_2$ and $\alpha_1$
are the bond angles between $d_1$ and $d_2$, and the two $d_1$ bonds respectively (see Fig.~\ref{structure}(b)).
Our calculated frequencies of $A^2_g$, $B_{2g}$ and $A^1_g$ modes using GGA functionals are 453 $cm^{-1}$, 424
$cm^{-1}$ and 354 $cm^{-1}$, which agree well with the observed frequencies (460 $cm^{-1}$, 438 $cm^{-1}$ and 360 $cm^{-1}$) of
these modes.

We now examine the minimum in the FWHM (at P$_{c1}$) which is related to the electron phonon coupling constant $\lambda$. 
The dependence of $\lambda$ of the three Raman modes on pressure shows a dip at -1.5 GPa. We have gone to negative pressure 
because GGA calculations show semiconductor to 
semi-metal transition  at P$_{c1}$ = -1.5 GPa.  The discrepancy in P$_{c1}$ using GGA functionals as compared to experiment (P$_{c1}$ = 1.1 GPa) 
and HSE functionals (P$_{c1}$ = 1.2 GPa) points to the inadequacy of the GGA functionals in estimation of bandgap and predicting the transition pressure. 
The corrected transition pressure will be P$_{c1}$$^{corrected}$=-1.5+2.7=1.2 GPa as indicated in Fig.~\ref{bandstructure}(d). 
This correction is used in the pressure axis in plotting Fig.~\ref{Figure-Pressure evolution of FWHM} (c). Further, with increasing pressure, the gap 
closes in the vicinity of Z-point at P$^{corrected}$ = 1.7 GPa (Fig.~\ref{bandstructure}(d)). 
Fig.~\ref{Figure-Pressure evolution of FWHM} (c) 
reveals that the $\lambda$ does show a shallow minimum at P$_{c1}$$^{corrected}$ =1.2 GPa for the $A^{1}_{g}$ and $A^{2}_{g}$ modes.
clearly establishing the link between a dip in $\lambda$ and FWHM with the ETT.
We did not see any anomaly in the $\lambda$ of $B_{2g}$ mode with pressure.  

The linewidth of a phonon mode of a crystal exhibiting a strong electron-phonon interaction is
largely determined by electron-phonon coupling (EPC). The relation between the EPC ($\lambda_{q\nu}$) and
FWHM ($\gamma_{q\nu}$) \cite{Pisana, Lazzeri} of a phonon $\nu$ at wavevector q is given by the following expression,
\begin{equation}
\lambda_{q\nu} = \frac{\gamma_{q\nu}}{\pi\hbar N(\epsilon_F) {\omega_{q\nu}}^{2}},
\end{equation}
where $N(\epsilon_F)$ is the density of electronic states at Fermi level. From Eq. (2), it is clear that $\lambda$ and 
FWHM are proportional to each other. Since the density of states and electronic bands involved in the matrix elements
$\gamma_{q\nu}$ change upon band inversion at the ETT, there is thus a causal relationship with anomaly in FWHM.
Hence, we see a minimum in FWHM and $\lambda$ of a mode at the transition pressure. The anomalous change
in observed FWHM may not in general indicate an electronic topological transition (ETT). In the present context, however, 
a careful analysis of calculated electronic structure of black phosphorous reveals a band inversion across the 
transition pressure ($\sim$ 1.2 GPa), indicating the occurrence of an ETT. Our calculations of $Z_2$ invariant 
across $P_{c1}$ $\sim$ 1.2 GPa confirm this topological transition from trivial
semiconductor to topological semimetallic state. ETT is accompanied by the Lifshitz transition, the electronic 
structure changes and the charge density of P atoms redistributes, leading to the anomaly of Raman spectra.

\begin{figure}   
\includegraphics[width=0.5\textwidth]{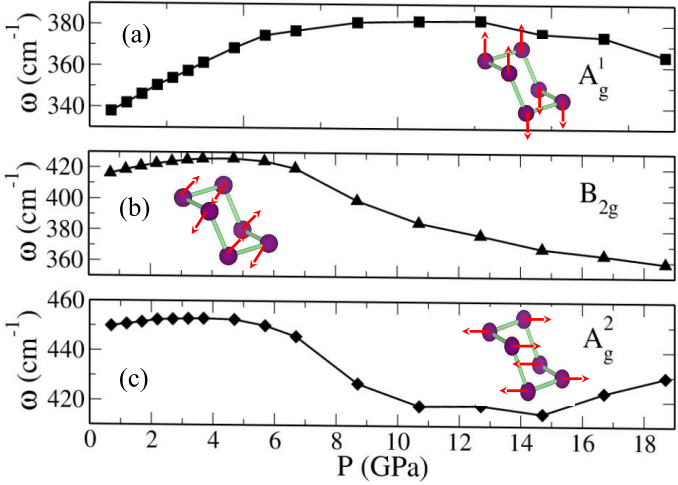}
\caption{Change in the calculated frequencies of Raman active modes (a) $A^{1}_{g}$, (b) $B_{2g}$ and (c) $A^{2}_{g}$ modes of BP 
in orthorhombic phase with pressure. Insets of (a), (b) and (c) show atomic displacements in these Raman active modes. The pressure 
axis has been corrected as explained in the text.}
\label{theoryphonon}
\end{figure}

\begin{figure}
\includegraphics[width=0.5\textwidth]{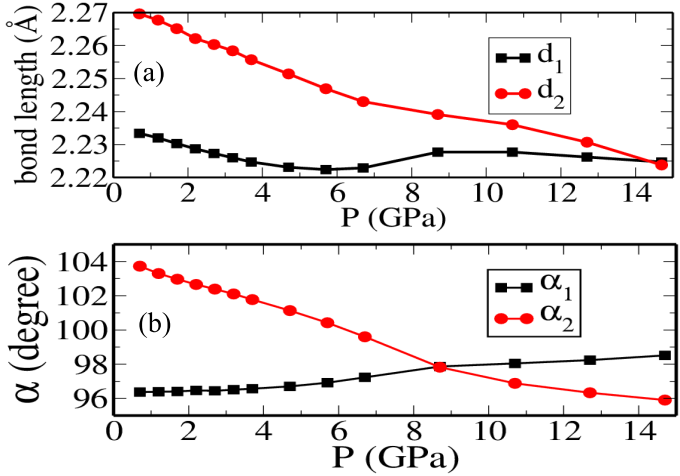}
\caption{Evolution of bond lengths (a) and bond angles (b) of BP in orthorhombic phase with pressure. The pressure axis has been corrected as 
explained in the text.}
\label{bondlength}
\end{figure}

\begin{figure} 
\includegraphics[width=0.5\textwidth]{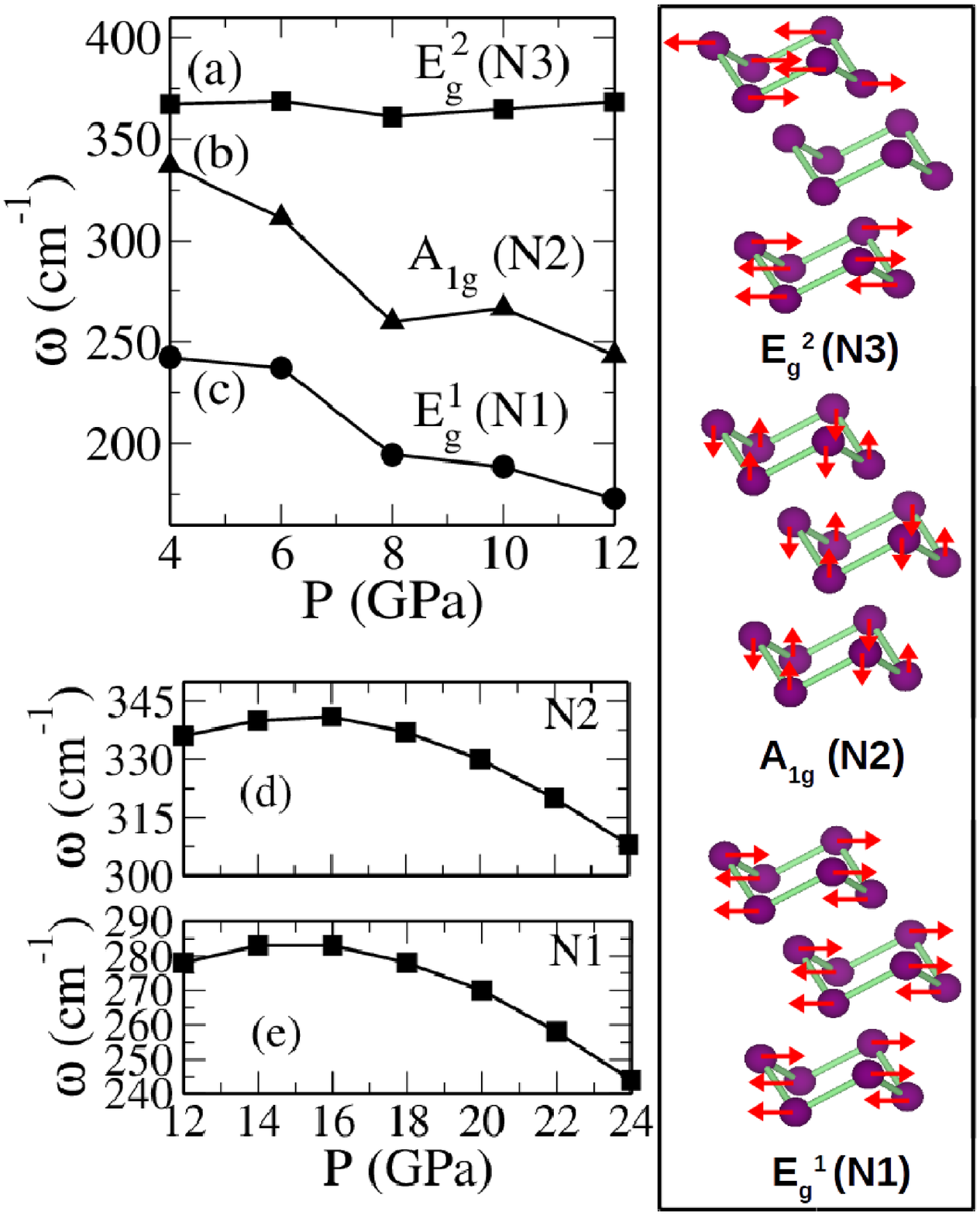}
\caption{Changes in the calculated frequencies of Raman active modes of BP, (a) $E_{g}^{2}$ (N3), (b) $A_{1g}$ (N2) 
and (c) $E_{g}^{1}$ (N1) modes in rhombohedral phase (right panel shows their atomic displacements) with pressure, (d) N2 and (e) N1 modes 
with pressure of BP in cubic phase. Note that N1, N2 and N3 are the modes observed in experiments. 
N2 and N1 modes as shown in figure (d) and (e) correspond to the acoustic modes at X and M-points in the 
Brillouin Zone of the simple cubic phase. 
\label{theoryA7phase}}
\end{figure}

\subsubsection{Anomalies at High Pressure Transitions: Structural Origin}
We now  address the pressure dependence of phonon frequencies of $A^{1}_{g}$, $A^{2}_{g}$ and $B_{2g}$ modes of the orthorhombic phase. The 
calculated frequencies of the orthorhombic phase as a function of pressure (Fig.~\ref{theoryphonon}) clearly 
display that the frequencies of $A^{1}_{g}$ and $A^{2}_{g}$ modes harden up to $\sim$ 5 GPa. Interestingly, the frequency 
of $B_{2g}$ and $A^{2}_{g}$ modes show softening at high pressure; resembling with the anomalous pressure dependence of the $B_{2g}$ 
and $A^{2}_{g}$ modes seen in the experiments between 7 and 9 GPa (Fig.~\ref{Figure-Pressure dependence of phonon frequencies}). 
To correlate the softening of the  $B_{2g}$ and $A^{2}_{g}$ modes with structural modifications, we examine the pressure 
dependence of the internal parameters $d_1$, $d_2$, $\alpha_1$ and $\alpha_2$ (see Fig. \ref{structure} for description).

Atomic vibrations of the $A^{1}_{g}$ mode (Fig.~\ref{theoryphonon}) involve stretching of the $d_2$ bond. From the 
variation in $d_2$ bond with pressure (Fig.~\ref{bondlength}(a)), we find that $d_2$ decreases monotonically with 
pressure and the $\Delta d_2$ is $\sim$ 0.04 {\AA} up to 14.7 GPa. As a result, $A^{1}_{g}$ hardens with pressure. 
The atomic vibrations of the $B_{2g}$ mode ( Fig.~\ref{theoryphonon}) affects bond angle $\alpha_2$, which decreases 
with increasing pressure (Fig.~\ref{bondlength}(b)), similar to  the frequency of $B_{2g}$ mode. 
Atomic displacements of the $A^{2}_{g}$ mode (Fig.~\ref{theoryphonon}) involve stretching of the $d_1$ 
bond. Fig.~\ref{bondlength}(a) shows that $d_1$ decreases with pressure upto 6 GPa, and starts increasing 
from 7-11 GPa, and subsequently decreases with pressure upto 14.7 GPa. In this case, the change in $d_1$ with pressure 
is much smaller  than that in $d_2$, which is also reflected in softening of $B_{2g}$ and $A^{2}_{g}$ modes. As a result 
of variation of $d_1$ with pressure, the $A^{2}_{g}$ mode softens with pressure and this softening is less than the 
softening of the $B_{2g}$ mode. The pressure axes of Fig.~\ref{theoryphonon} and  Fig.~\ref{bondlength} are corrected with a shift of +2.7 GPa as 
described in the previous section. Our theoretical calculations reveal that black phosphorous is 
quite a soft material, with its cell volume at 14.7 GPa being 24\%
smaller than that at 0 GPa. This large change in volume with pressure is responsible for significant softening of
$B_{2g}$ and $A^{2}_{g}$ modes. For such large changes in volume, we expect anharmonic effects to be strong, which are not
included in our analysis. This is probably responsible for some discrepancy between theory and experiments.
We find that the observed softening of $A^{2}_{g}$ mode (Fig.~\ref{theoryphonon}) is not indefinite:
$A^{2}_{g}$ mode harden with further increase in pressure beyond 15 GPa, consistent with our experiments.

Orthorhombic phase is a low pressure phase of black phosphorous which is stable only at P $<$ 5
GPa. We presented results of our calculations on orthorhombic phase at P $>$ 5 GPa because experiments are indicative 
of the presence of mixed phases at higher pressure, as evidenced in the survival of $A_g^1$, $B_{2g}$  and $A_g^2$ modes 
 at high pressures along with the modes N1 ($E_g$), 
N2 ($A_{1g}$) and N3 ($E_g$) of the rhombohedral phase (refer Fig.~\ref{Figure-Pressure dependence of phonon frequencies}). As the calculations here are on pure and not on mixed phases 
(orthorhombic and rhombohedral phases together as seen in experiment), interpretation of experimental and theoretical 
results should be done with care.

The space group of BP in rhombohedral structure (also known as A7 phase after 4.6 GPa) is $R\bar{3}m$ with 
primitive cell containing 2 atoms. Group theory gives zone center  optical modes as A$_{1g}$+E$_{g}$+A$_{2u}$+E$_{u}$, 
where $A_{1g}$ and $E_{g}$ are the Raman active modes and $A_{2u}$ and $E_{u}$ are infrared active modes. 
We used hexagonal unit cell with 6 atoms (\textit{i.e.} a supercell) to perform phonon calculations 
for A7 phase. Some of the zone boundary points of BZ of rhombohedral structure (\textit{i.e.} the primitive cell) 
fold onto the $\Gamma$-point of BZ of the hexagonal 
unit cell, and enable the analysis of Raman active modes and their symmetries. The frequencies of the Raman active modes 
$E_g^2$, $A_{1g}$ and $E_g^1$ modes are 367 $cm^{-1}$, 336 cm$^{-1}$ and 242 cm$^{-1}$ respectively at 4 GPa. Calculated frequencies 
of these modes are in good agreement with experimental values (\textit{e.g.} at 6 GPa, $\omega$ (N1) = 237cm$^{-1}$ 
(Exp. 237cm$^{-1}$), $\omega$ (N2) = 311 cm$^{-1}$ (Exp. 307 cm$^{-1}$)). We identify the experimentally observed N3 mode as $E_g^2$, N2 
mode as $A_{1g}$ mode and N1 mode as $E_g^1$. 
Symmetry assignment of N2 and N1 modes is consistent with previous report \cite{akahama1997raman}. All the three modes N1, N2  and 
N3 soften with pressure, in agreement with the observed trends in our experiment 
(see Fig.~\ref{theoryA7phase} and Fig.~\ref{Figure-Pressure dependence of phonon frequencies}).
From first-principles calculations, we find that the A7-phase (hexagonal structure) is stable at pressures 
from 4 GPa to 12 GPa and, as unstable modes appear in phonon dispersion beyond this pressure. The N1 and N2 modes also persist at 
pressures till 24 GPa (sc phase). It is likely that the A7-phase exists beyond 11 GPa. There are no Raman active modes in the sc 
phase with one atom per unit cell. 
Another possibility is that the modes N1 and N2 are the disorder activated zone boundary acoustic phonons of the sc phase. 
To examine this possibility, we carried out  calculations of the zone boundary acoustic phonons of the sc phase. 
Our calculations indeed show acoustic modes at X point (283 cm$^{-1}$) and M point 
(340 cm$^{-1}$), which exhibit the pressure dependence similar to that of  N1 and N2 modes \textit{i.e} 
 both of them soften with pressure (Fig.~\ref{theoryA7phase}). 
Thus, we have assigned N1 and N2 modes in the sc phase to these zone boundary acoustic modes.

\section{Conclusion}
High pressure Raman study of BP up to 24 GPa is reported using diamond anvil cell under hydrostatic conditions. 
The line width of all the three  Raman modes A$^1_g$, B$_{2g}$ and A$^2_g$ of the orthorhombic phase decrease 
with pressure up to 1.1 GPa and then increase, thus showing a minimum at $\sim$ 1.1 GPa corresponding to the 
transition pressure of ETT. Using first-principles calculations, we find that the low pressure phase transition 
in BP  is a semiconductor to topological insulator transition. Our calculations of the $Z_2$ invariant confirm the 
change in electronic topology marking a transition from band to topological insulating state. The Raman shift of 
these modes show normal hardening with pressure in orthorhombic phase. B$_{2g}$ and A$^2_g$ modes show anomalous 
softening in rhombohedral phase at 7.4 GPa while A$^1_g$ mode exhibits normal hardening. These modes disappear 
completely at 11 GPa in the sc phase. The new modes N1, N2 and N3 arising in rhombohedral phase also show anomalous 
softening of the frequency with pressure. We uncover the origin of these anomalous softening of Raman  modes in 
the variation of internal structural parameters with pressure, and identify the symmetry of the new modes appearing 
upon pressure in experiments.

\section{Acknowledgements}
AS is greatful to Jawaharlal Nehru Centre for Scientific Advanced Research, India for research fellowship.
AKS and UVW thank DST for the JC Bose National fellowships. AKS thanks DST Nanomission for the financial support.

\clearpage
%\bibliography{ref}

\begin{thebibliography}{50}
\expandafter\ifx\csname natexlab\endcsname\relax\def\natexlab#1{#1}\fi
\expandafter\ifx\csname bibnamefont\endcsname\relax
  \def\bibnamefont#1{#1}\fi
\expandafter\ifx\csname bibfnamefont\endcsname\relax
  \def\bibfnamefont#1{#1}\fi
\expandafter\ifx\csname citenamefont\endcsname\relax
  \def\citenamefont#1{#1}\fi
\expandafter\ifx\csname url\endcsname\relax
  \def\url#1{\texttt{#1}}\fi
\expandafter\ifx\csname urlprefix\endcsname\relax\def\urlprefix{URL }\fi
\providecommand{\bibinfo}[2]{#2}
\providecommand{\eprint}[2][]{\url{#2}}

\bibitem[{\citenamefont{Ling et~al.}(2015)\citenamefont{Ling, Wang, Huang, Xia,
  and Dresselhaus}}]{ling2015renaissance}
\bibinfo{author}{\bibfnamefont{X.}~\bibnamefont{Ling}},
  \bibinfo{author}{\bibfnamefont{H.}~\bibnamefont{Wang}},
  \bibinfo{author}{\bibfnamefont{S.}~\bibnamefont{Huang}},
  \bibinfo{author}{\bibfnamefont{F.}~\bibnamefont{Xia}}, \bibnamefont{and}
  \bibinfo{author}{\bibfnamefont{M.~S.} \bibnamefont{Dresselhaus}},
  \bibinfo{journal}{Proc. Natl. Acad. Sci.} \textbf{\bibinfo{volume}{112}},
  \bibinfo{pages}{4523} (\bibinfo{year}{2015}).

\bibitem[{\citenamefont{Chakraborty et~al.}(2016)\citenamefont{Chakraborty,
  Gupta, Singh, Kuiri, Kumar, Muthu, Das, Waghmare, and
  Sood}}]{chakraborty2016electron}
\bibinfo{author}{\bibfnamefont{B.}~\bibnamefont{Chakraborty}},
  \bibinfo{author}{\bibfnamefont{S.~N.} \bibnamefont{Gupta}},
  \bibinfo{author}{\bibfnamefont{A.}~\bibnamefont{Singh}},
  \bibinfo{author}{\bibfnamefont{M.}~\bibnamefont{Kuiri}},
  \bibinfo{author}{\bibfnamefont{C.}~\bibnamefont{Kumar}},
  \bibinfo{author}{\bibfnamefont{D.}~\bibnamefont{Muthu}},
  \bibinfo{author}{\bibfnamefont{A.}~\bibnamefont{Das}},
  \bibinfo{author}{\bibfnamefont{U.}~\bibnamefont{Waghmare}}, \bibnamefont{and}
  \bibinfo{author}{\bibfnamefont{A.}~\bibnamefont{Sood}}, \bibinfo{journal}{2D
  Materials} \textbf{\bibinfo{volume}{3}}, \bibinfo{pages}{015008}
  (\bibinfo{year}{2016}).

\bibitem[{\citenamefont{Li et~al.}(2014)\citenamefont{Li, Yu, Ye, Ge, Ou, Wu,
  Feng, Chen, and Zhang}}]{li2014black}
\bibinfo{author}{\bibfnamefont{L.}~\bibnamefont{Li}},
  \bibinfo{author}{\bibfnamefont{Y.}~\bibnamefont{Yu}},
  \bibinfo{author}{\bibfnamefont{G.~J.} \bibnamefont{Ye}},
  \bibinfo{author}{\bibfnamefont{Q.}~\bibnamefont{Ge}},
  \bibinfo{author}{\bibfnamefont{X.}~\bibnamefont{Ou}},
  \bibinfo{author}{\bibfnamefont{H.}~\bibnamefont{Wu}},
  \bibinfo{author}{\bibfnamefont{D.}~\bibnamefont{Feng}},
  \bibinfo{author}{\bibfnamefont{X.~H.} \bibnamefont{Chen}}, \bibnamefont{and}
  \bibinfo{author}{\bibfnamefont{Y.}~\bibnamefont{Zhang}},
  \bibinfo{journal}{Nat. Nanotechnol.} \textbf{\bibinfo{volume}{9}},
  \bibinfo{pages}{372} (\bibinfo{year}{2014}).

\bibitem[{\citenamefont{Xia et~al.}(2014)\citenamefont{Xia, Wang, and
  Jia}}]{xia2014rediscovering}
\bibinfo{author}{\bibfnamefont{F.}~\bibnamefont{Xia}},
  \bibinfo{author}{\bibfnamefont{H.}~\bibnamefont{Wang}}, \bibnamefont{and}
  \bibinfo{author}{\bibfnamefont{Y.}~\bibnamefont{Jia}},
  \bibinfo{journal}{Nature commun.} \textbf{\bibinfo{volume}{5}},\bibinfo{pages}{4458}
  (\bibinfo{year}{2014}).

\bibitem[{\citenamefont{Liu et~al.}(2014)\citenamefont{Liu, Neal, Zhu, Luo, Xu,
  Tom{\'a}nek, and Ye}}]{liu2014phosphorene}
\bibinfo{author}{\bibfnamefont{H.}~\bibnamefont{Liu}},
  \bibinfo{author}{\bibfnamefont{A.~T.} \bibnamefont{Neal}},
  \bibinfo{author}{\bibfnamefont{Z.}~\bibnamefont{Zhu}},
  \bibinfo{author}{\bibfnamefont{Z.}~\bibnamefont{Luo}},
  \bibinfo{author}{\bibfnamefont{X.}~\bibnamefont{Xu}},
  \bibinfo{author}{\bibfnamefont{D.}~\bibnamefont{Tom{\'a}nek}},
  \bibnamefont{and} \bibinfo{author}{\bibfnamefont{P.~D.} \bibnamefont{Ye}},
  \bibinfo{journal}{ACS Nano} \textbf{\bibinfo{volume}{8}},
  \bibinfo{pages}{4033} (\bibinfo{year}{2014}).

\bibitem[{\citenamefont{Koenig et~al.}(2014)\citenamefont{Koenig, Doganov,
  Schmidt, Neto, and Oezyilmaz}}]{koenig2014electric}
\bibinfo{author}{\bibfnamefont{S.~P.} \bibnamefont{Koenig}},
  \bibinfo{author}{\bibfnamefont{R.~A.} \bibnamefont{Doganov}},
  \bibinfo{author}{\bibfnamefont{H.}~\bibnamefont{Schmidt}},
  \bibinfo{author}{\bibfnamefont{A.~C.} \bibnamefont{Neto}}, \bibnamefont{and}
  \bibinfo{author}{\bibfnamefont{B.}~\bibnamefont{Oezyilmaz}},
  \bibinfo{journal}{App. Phys. Lett.} \textbf{\bibinfo{volume}{104}},
  \bibinfo{pages}{103106} (\bibinfo{year}{2014}).

\bibitem[{\citenamefont{Rodin et~al.}(2014)\citenamefont{Rodin, Carvalho, and
  Neto}}]{rodin2014strain}
\bibinfo{author}{\bibfnamefont{A.}~\bibnamefont{Rodin}},
  \bibinfo{author}{\bibfnamefont{A.}~\bibnamefont{Carvalho}}, \bibnamefont{and}
  \bibinfo{author}{\bibfnamefont{A.~C.} \bibnamefont{Neto}},
  \bibinfo{journal}{Phys. Rev. lett.} \textbf{\bibinfo{volume}{112}},
  \bibinfo{pages}{176801} (\bibinfo{year}{2014}).

\bibitem[{\citenamefont{Qiao et~al.}(2014)\citenamefont{Qiao, Kong, Hu, Yang,
  and Ji}}]{qiao2014high}
\bibinfo{author}{\bibfnamefont{J.}~\bibnamefont{Qiao}},
  \bibinfo{author}{\bibfnamefont{X.}~\bibnamefont{Kong}},
  \bibinfo{author}{\bibfnamefont{Z.-X.} \bibnamefont{Hu}},
  \bibinfo{author}{\bibfnamefont{F.}~\bibnamefont{Yang}}, \bibnamefont{and}
  \bibinfo{author}{\bibfnamefont{W.}~\bibnamefont{Ji}},
  \bibinfo{journal}{Nature Commun.} \textbf{\bibinfo{volume}{5}},\bibinfo{pages}{4475}
  (\bibinfo{year}{2014}).

\bibitem[{\citenamefont{Zhang et~al.}(2014)\citenamefont{Zhang, Yang, Xu, Wang,
  Li, Ghufran, Zhang, Yu, Zhang, Qin et~al.}}]{zhang2014extraordinary}
\bibinfo{author}{\bibfnamefont{S.}~\bibnamefont{Zhang}},
  \bibinfo{author}{\bibfnamefont{J.}~\bibnamefont{Yang}},
  \bibinfo{author}{\bibfnamefont{R.}~\bibnamefont{Xu}},
  \bibinfo{author}{\bibfnamefont{F.}~\bibnamefont{Wang}},
  \bibinfo{author}{\bibfnamefont{W.}~\bibnamefont{Li}},
  \bibinfo{author}{\bibfnamefont{M.}~\bibnamefont{Ghufran}},
  \bibinfo{author}{\bibfnamefont{Y.-W.} \bibnamefont{Zhang}},
  \bibinfo{author}{\bibfnamefont{Z.}~\bibnamefont{Yu}},
  \bibinfo{author}{\bibfnamefont{G.}~\bibnamefont{Zhang}},
  \bibinfo{author}{\bibfnamefont{Q.}~\bibnamefont{Qin}}, \bibnamefont{et~al.},
  \bibinfo{journal}{ACS Nano} \textbf{\bibinfo{volume}{8}},
  \bibinfo{pages}{9590} (\bibinfo{year}{2014}).

\bibitem[{\citenamefont{Novoselov et~al.}(2005)\citenamefont{Novoselov, Geim,
  Morozov, Jiang, Katsnelson, Grigorieva, Dubonos, and
  Firsov}}]{novoselov2005two}
\bibinfo{author}{\bibfnamefont{K.}~\bibnamefont{Novoselov}},
  \bibinfo{author}{\bibfnamefont{A.~K.} \bibnamefont{Geim}},
  \bibinfo{author}{\bibfnamefont{S.}~\bibnamefont{Morozov}},
  \bibinfo{author}{\bibfnamefont{D.}~\bibnamefont{Jiang}},
  \bibinfo{author}{\bibfnamefont{M.}~\bibnamefont{Katsnelson}},
  \bibinfo{author}{\bibfnamefont{I.}~\bibnamefont{Grigorieva}},
  \bibinfo{author}{\bibfnamefont{S.}~\bibnamefont{Dubonos}}, \bibnamefont{and}
  \bibinfo{author}{\bibfnamefont{A.}~\bibnamefont{Firsov}},
  \bibinfo{journal}{Nature} \textbf{\bibinfo{volume}{438}},
  \bibinfo{pages}{197} (\bibinfo{year}{2005}).

\bibitem[{\citenamefont{Zhang et~al.}(2005)\citenamefont{Zhang, Tan, Stormer,
  and Kim}}]{zhang2005experimental}
\bibinfo{author}{\bibfnamefont{Y.}~\bibnamefont{Zhang}},
  \bibinfo{author}{\bibfnamefont{Y.-W.} \bibnamefont{Tan}},
  \bibinfo{author}{\bibfnamefont{H.~L.} \bibnamefont{Stormer}},
  \bibnamefont{and} \bibinfo{author}{\bibfnamefont{P.}~\bibnamefont{Kim}},
  \bibinfo{journal}{Nature} \textbf{\bibinfo{volume}{438}},
  \bibinfo{pages}{201} (\bibinfo{year}{2005}).

\bibitem[{\citenamefont{Mak et~al.}(2010)\citenamefont{Mak, Lee, Hone, Shan,
  and Heinz}}]{mak2010atomically}
\bibinfo{author}{\bibfnamefont{K.~F.} \bibnamefont{Mak}},
  \bibinfo{author}{\bibfnamefont{C.}~\bibnamefont{Lee}},
  \bibinfo{author}{\bibfnamefont{J.}~\bibnamefont{Hone}},
  \bibinfo{author}{\bibfnamefont{J.}~\bibnamefont{Shan}}, \bibnamefont{and}
  \bibinfo{author}{\bibfnamefont{T.~F.} \bibnamefont{Heinz}},
  \bibinfo{journal}{Phys. Rev. lett.} \textbf{\bibinfo{volume}{105}},
  \bibinfo{pages}{136805} (\bibinfo{year}{2010}).

\bibitem[{\citenamefont{Splendiani et~al.}(2010)\citenamefont{Splendiani, Sun,
  Zhang, Li, Kim, Chim, Galli, and Wang}}]{splendiani2010emerging}
\bibinfo{author}{\bibfnamefont{A.}~\bibnamefont{Splendiani}},
  \bibinfo{author}{\bibfnamefont{L.}~\bibnamefont{Sun}},
  \bibinfo{author}{\bibfnamefont{Y.}~\bibnamefont{Zhang}},
  \bibinfo{author}{\bibfnamefont{T.}~\bibnamefont{Li}},
  \bibinfo{author}{\bibfnamefont{J.}~\bibnamefont{Kim}},
  \bibinfo{author}{\bibfnamefont{C.-Y.} \bibnamefont{Chim}},
  \bibinfo{author}{\bibfnamefont{G.}~\bibnamefont{Galli}}, \bibnamefont{and}
  \bibinfo{author}{\bibfnamefont{F.}~\bibnamefont{Wang}},
  \bibinfo{journal}{Nano Lett.} \textbf{\bibinfo{volume}{10}},
  \bibinfo{pages}{1271} (\bibinfo{year}{2010}).

\bibitem[{\citenamefont{Jones et~al.}(2014)\citenamefont{Jones, Yu, Ross,
  Klement, Ghimire, Yan, Mandrus, Yao, and Xu}}]{jones2014spin}
\bibinfo{author}{\bibfnamefont{A.~M.} \bibnamefont{Jones}},
  \bibinfo{author}{\bibfnamefont{H.}~\bibnamefont{Yu}},
  \bibinfo{author}{\bibfnamefont{J.~S.} \bibnamefont{Ross}},
  \bibinfo{author}{\bibfnamefont{P.}~\bibnamefont{Klement}},
  \bibinfo{author}{\bibfnamefont{N.~J.} \bibnamefont{Ghimire}},
  \bibinfo{author}{\bibfnamefont{J.}~\bibnamefont{Yan}},
  \bibinfo{author}{\bibfnamefont{D.~G.} \bibnamefont{Mandrus}},
  \bibinfo{author}{\bibfnamefont{W.}~\bibnamefont{Yao}}, \bibnamefont{and}
  \bibinfo{author}{\bibfnamefont{X.}~\bibnamefont{Xu}},
  \bibinfo{journal}{Nature Phys.} \textbf{\bibinfo{volume}{10}},
  \bibinfo{pages}{130} (\bibinfo{year}{2014}).

\bibitem[{\citenamefont{Liu et~al.}(2015)\citenamefont{Liu, Du, Deng, and
  Peide}}]{liu2015semiconducting}
\bibinfo{author}{\bibfnamefont{H.}~\bibnamefont{Liu}},
  \bibinfo{author}{\bibfnamefont{Y.}~\bibnamefont{Du}},
  \bibinfo{author}{\bibfnamefont{Y.}~\bibnamefont{Deng}}, \bibnamefont{and}
  \bibinfo{author}{\bibfnamefont{D.~Y.} \bibnamefont{Peide}},
  \bibinfo{journal}{Chem. Soc. Rev.} \textbf{\bibinfo{volume}{44}},
  \bibinfo{pages}{2732} (\bibinfo{year}{2015}).

\bibitem[{\citenamefont{Jamieson}(1963)}]{jamieson1963crystal}
\bibinfo{author}{\bibfnamefont{J.~C.} \bibnamefont{Jamieson}},
  \bibinfo{journal}{Science} \textbf{\bibinfo{volume}{139}},
  \bibinfo{pages}{1291} (\bibinfo{year}{1963}).

\bibitem[{\citenamefont{Cartz et~al.}(1979)\citenamefont{Cartz, Srinivasa,
  Riedner, Jorgensen, and Worlton}}]{cartz1979effect}
\bibinfo{author}{\bibfnamefont{L.}~\bibnamefont{Cartz}},
  \bibinfo{author}{\bibfnamefont{S.}~\bibnamefont{Srinivasa}},
  \bibinfo{author}{\bibfnamefont{R.}~\bibnamefont{Riedner}},
  \bibinfo{author}{\bibfnamefont{J.}~\bibnamefont{Jorgensen}},
  \bibnamefont{and} \bibinfo{author}{\bibfnamefont{T.}~\bibnamefont{Worlton}},
  \bibinfo{journal}{J. Chem. Phys.} \textbf{\bibinfo{volume}{71}},
  \bibinfo{pages}{1718} (\bibinfo{year}{1979}).

\bibitem[{\citenamefont{Kikegawa and Iwasaki}(1983)}]{kikegawa1983x}
\bibinfo{author}{\bibfnamefont{T.}~\bibnamefont{Kikegawa}} \bibnamefont{and}
  \bibinfo{author}{\bibfnamefont{H.}~\bibnamefont{Iwasaki}},
  \bibinfo{journal}{Acta Crystallogr. Sect. B} \textbf{\bibinfo{volume}{39}},
  \bibinfo{pages}{158} (\bibinfo{year}{1983}).

\bibitem[{\citenamefont{Burdett and Lee}(1982)}]{burdett1982pressure}
\bibinfo{author}{\bibfnamefont{J.~K.} \bibnamefont{Burdett}} \bibnamefont{and}
  \bibinfo{author}{\bibfnamefont{S.}~\bibnamefont{Lee}}, \bibinfo{journal}{J.
  Solid State Chem.} \textbf{\bibinfo{volume}{44}}, \bibinfo{pages}{415}
  (\bibinfo{year}{1982}).

\bibitem[{\citenamefont{Sugai et~al.}(1981)\citenamefont{Sugai, Ueda, and
  Murase}}]{sugai1981pressure}
\bibinfo{author}{\bibfnamefont{S.}~\bibnamefont{Sugai}},
  \bibinfo{author}{\bibfnamefont{T.}~\bibnamefont{Ueda}}, \bibnamefont{and}
  \bibinfo{author}{\bibfnamefont{K.}~\bibnamefont{Murase}},
  \bibinfo{journal}{J. Phys. Soc. Jpn.} \textbf{\bibinfo{volume}{50}},
  \bibinfo{pages}{3356} (\bibinfo{year}{1981}).

\bibitem[{\citenamefont{Vanderborgh and Schiferl}(1989)}]{vanderborgh1989raman}
\bibinfo{author}{\bibfnamefont{C.}~\bibnamefont{Vanderborgh}} \bibnamefont{and}
  \bibinfo{author}{\bibfnamefont{D.}~\bibnamefont{Schiferl}},
  \bibinfo{journal}{Phys. Rev. B} \textbf{\bibinfo{volume}{40}},
  \bibinfo{pages}{9595} (\bibinfo{year}{1989}).

\bibitem[{\citenamefont{Akahama et~al.}(1997)\citenamefont{Akahama, Kobayashi,
  and Kawamura}}]{akahama1997raman}
\bibinfo{author}{\bibfnamefont{Y.}~\bibnamefont{Akahama}},
  \bibinfo{author}{\bibfnamefont{M.}~\bibnamefont{Kobayashi}},
  \bibnamefont{and} \bibinfo{author}{\bibfnamefont{H.}~\bibnamefont{Kawamura}},
  \bibinfo{journal}{Solid State Commun.} \textbf{\bibinfo{volume}{104}},
  \bibinfo{pages}{311} (\bibinfo{year}{1997}).

\bibitem[{\citenamefont{Akahama et~al.}(1999)\citenamefont{Akahama, Kobayashi,
  and Kawamura}}]{akahama1999simple}
\bibinfo{author}{\bibfnamefont{Y.}~\bibnamefont{Akahama}},
  \bibinfo{author}{\bibfnamefont{M.}~\bibnamefont{Kobayashi}},
  \bibnamefont{and} \bibinfo{author}{\bibfnamefont{H.}~\bibnamefont{Kawamura}},
  \bibinfo{journal}{Phys. Rev. B} \textbf{\bibinfo{volume}{59}},
  \bibinfo{pages}{8520} (\bibinfo{year}{1999}).

\bibitem[{\citenamefont{Akahama et~al.}(2000)\citenamefont{Akahama, Kawamura,
  Carlson, Le~Bihan, and H{\"a}usermann}}]{akahama2000structural}
\bibinfo{author}{\bibfnamefont{Y.}~\bibnamefont{Akahama}},
  \bibinfo{author}{\bibfnamefont{H.}~\bibnamefont{Kawamura}},
  \bibinfo{author}{\bibfnamefont{S.}~\bibnamefont{Carlson}},
  \bibinfo{author}{\bibfnamefont{T.}~\bibnamefont{Le~Bihan}}, \bibnamefont{and}
  \bibinfo{author}{\bibfnamefont{D.}~\bibnamefont{H{\"a}usermann}},
  \bibinfo{journal}{Phys. Rev. B} \textbf{\bibinfo{volume}{61}},
  \bibinfo{pages}{3139} (\bibinfo{year}{2000}).

\bibitem[{\citenamefont{Wittig and Matthias}(1968)}]{wittig1968superconducting}
\bibinfo{author}{\bibfnamefont{J.}~\bibnamefont{Wittig}} \bibnamefont{and}
  \bibinfo{author}{\bibfnamefont{B.}~\bibnamefont{Matthias}},
  \bibinfo{journal}{Science} \textbf{\bibinfo{volume}{160}},
  \bibinfo{pages}{994} (\bibinfo{year}{1968}).

\bibitem[{\citenamefont{Kawamura et~al.}(1984)\citenamefont{Kawamura,
  Shirotani, and Tachikawa}}]{kawamura1984anomalous}
\bibinfo{author}{\bibfnamefont{H.}~\bibnamefont{Kawamura}},
  \bibinfo{author}{\bibfnamefont{I.}~\bibnamefont{Shirotani}},
  \bibnamefont{and}
  \bibinfo{author}{\bibfnamefont{K.}~\bibnamefont{Tachikawa}},
  \bibinfo{journal}{Solid state commun.} \textbf{\bibinfo{volume}{49}},
  \bibinfo{pages}{879} (\bibinfo{year}{1984}).




\bibitem{Akahama 2001} Y. Akahama and H. Kawamura, Phys. Status Solidi B \textbf{223}, 349 (2001).



\bibitem{Okajima 1984} M. Okajima, S. Endo, Y. Akahama and S. Narita, Jpn. J. Appl. Phys. \textbf{23}, 15, (1984).

\bibitem[{\citenamefont{Akahama et~al.}(1986)\citenamefont{Akahama, Endo, and
  Narita}}]{akahama1986electrical}
\bibinfo{author}{\bibfnamefont{Y.}~\bibnamefont{Akahama}},
  \bibinfo{author}{\bibfnamefont{S.}~\bibnamefont{Endo}}, \bibnamefont{and}
  \bibinfo{author}{\bibfnamefont{S.}~\bibnamefont{Narita}},
  \bibinfo{journal}{Physica B+ C} \textbf{\bibinfo{volume}{139}},
  \bibinfo{pages}{397} (\bibinfo{year}{1986}).

\bibitem[{\citenamefont{Gong et~al.}(2015)\citenamefont{Gong, Liu, Yan, Xiang,
  Chen, Shen, and Zou}}]{gong2015hydrostatic}
\bibinfo{author}{\bibfnamefont{P.-L.} \bibnamefont{Gong}},
  \bibinfo{author}{\bibfnamefont{D.-Y.} \bibnamefont{Liu}},
  \bibinfo{author}{\bibfnamefont{K.-S.} \bibnamefont{Yan}},
  \bibinfo{author}{\bibfnamefont{Z.-J.} \bibnamefont{Xiang}},
  \bibinfo{author}{\bibfnamefont{X.-H.} \bibnamefont{Chen}},
  \bibinfo{author}{\bibfnamefont{S.-Q.} \bibnamefont{Shen}}, \bibnamefont{and}
  \bibinfo{author}{\bibfnamefont{L.-J.} \bibnamefont{Zou}},
   \bibinfo{journal}{Phys. Rev. B} \textbf{\bibinfo{volume}{93}},
    \bibinfo{pages}{195434} (\bibinfo{year}{2016}).



\bibitem[{\citenamefont{Xiang et~al.}(2015{\natexlab{a}})\citenamefont{Xiang,
  Ye, Shang, Lei, Wang, Yang, Liu, Meng, Luo, Zou
  et~al.}}]{PhysRevLett.115.186403}
\bibinfo{author}{\bibfnamefont{Z.~J.} \bibnamefont{Xiang}},
  \bibinfo{author}{\bibfnamefont{G.~J.} \bibnamefont{Ye}},
  \bibinfo{author}{\bibfnamefont{C.}~\bibnamefont{Shang}},
  \bibinfo{author}{\bibfnamefont{B.}~\bibnamefont{Lei}},
  \bibinfo{author}{\bibfnamefont{N.~Z.} \bibnamefont{Wang}},
  \bibinfo{author}{\bibfnamefont{K.~S.} \bibnamefont{Yang}},
  \bibinfo{author}{\bibfnamefont{D.~Y.} \bibnamefont{Liu}},
  \bibinfo{author}{\bibfnamefont{F.~B.} \bibnamefont{Meng}},
  \bibinfo{author}{\bibfnamefont{X.~G.} \bibnamefont{Luo}},
  \bibinfo{author}{\bibfnamefont{L.~J.} \bibnamefont{Zou}},
  \bibnamefont{et~al.}, \bibinfo{journal}{Phys. Rev. Lett.}
  \textbf{\bibinfo{volume}{115}}, \bibinfo{pages}{186403}
  (\bibinfo{year}{2015}{\natexlab{a}}).




\bibitem[{\citenamefont{Bera et~al.}(2013)\citenamefont{Bera, Pal, Muthu, Sen,
  Guptasarma, Waghmare, and Sood}}]{Bera}
\bibinfo{author}{\bibfnamefont{A.}~\bibnamefont{Bera}},
  \bibinfo{author}{\bibfnamefont{K.}~\bibnamefont{Pal}},
  \bibinfo{author}{\bibfnamefont{D.~V.~S.} \bibnamefont{Muthu}},
  \bibinfo{author}{\bibfnamefont{S.}~\bibnamefont{Sen}},
  \bibinfo{author}{\bibfnamefont{P.}~\bibnamefont{Guptasarma}},
  \bibinfo{author}{\bibfnamefont{U.~V.} \bibnamefont{Waghmare}},
  \bibnamefont{and} \bibinfo{author}{\bibfnamefont{A.~K.} \bibnamefont{Sood}},
  \bibinfo{journal}{Phys. Rev. Lett.} \textbf{\bibinfo{volume}{110}},
  \bibinfo{pages}{107401} (\bibinfo{year}{2013}).

\bibitem[{\citenamefont{Pal and Waghmare}(2014)}]{pal}
\bibinfo{author}{\bibfnamefont{K.}~\bibnamefont{Pal}} \bibnamefont{and}
  \bibinfo{author}{\bibfnamefont{U.~V.} \bibnamefont{Waghmare}},
  \bibinfo{journal}{Appl. Phys. Lett.} \textbf{\bibinfo{volume}{105}},
  \bibinfo{pages}{062105} (\bibinfo{year}{2014}).

\bibitem[{\citenamefont{Xi et~al.}(2013)\citenamefont{Xi, Ma, Liu, Chen, Ku,
  Berger, Martin, Tanner, and Carr}}]{xi}
\bibinfo{author}{\bibfnamefont{X.}~\bibnamefont{Xi}},
  \bibinfo{author}{\bibfnamefont{C.}~\bibnamefont{Ma}},
  \bibinfo{author}{\bibfnamefont{Z.}~\bibnamefont{Liu}},
  \bibinfo{author}{\bibfnamefont{Z.}~\bibnamefont{Chen}},
  \bibinfo{author}{\bibfnamefont{W.}~\bibnamefont{Ku}},
  \bibinfo{author}{\bibfnamefont{H.}~\bibnamefont{Berger}},
  \bibinfo{author}{\bibfnamefont{C.}~\bibnamefont{Martin}},
  \bibinfo{author}{\bibfnamefont{D.~B.} \bibnamefont{Tanner}},
  \bibnamefont{and} \bibinfo{author}{\bibfnamefont{G.~L.} \bibnamefont{Carr}},
  \bibinfo{journal}{Phys. Rev. Lett.} \textbf{\bibinfo{volume}{111}},
  \bibinfo{pages}{155701} (\bibinfo{year}{2013}).



\bibitem{Manjanath} A. Manjanath, A. Samanta, T. Pandey and A. K. Singh, Nanotechnology, \textbf{26}, 075701, (2015).


\bibitem{Akiba2015} K. Akiba, A. Miyake and Y. Akahama, K. Matsubayashi, Y. Uwatoko, H. Arai, Y. Fuseya and M. Tokunaga, J. Phys. Soc. Jpn, \textbf{84}, 073708, (2015).

\bibitem[{\citenamefont{Yamada et~al.}(1984)\citenamefont{Yamada, Fujii,
  Akahama, Endo, Narita, Axe, and McWhan}}]{PhysRevB.30.2410}
\bibinfo{author}{\bibfnamefont{Y.}~\bibnamefont{Yamada}},
  \bibinfo{author}{\bibfnamefont{Y.}~\bibnamefont{Fujii}},
  \bibinfo{author}{\bibfnamefont{Y.}~\bibnamefont{Akahama}},
  \bibinfo{author}{\bibfnamefont{S.}~\bibnamefont{Endo}},
  \bibinfo{author}{\bibfnamefont{S.}~\bibnamefont{Narita}},
  \bibinfo{author}{\bibfnamefont{J.~D.} \bibnamefont{Axe}}, \bibnamefont{and}
  \bibinfo{author}{\bibfnamefont{D.~B.} \bibnamefont{McWhan}},
  \bibinfo{journal}{Phys. Rev. B} \textbf{\bibinfo{volume}{30}},
  \bibinfo{pages}{2410} (\bibinfo{year}{1984}).
  
  
  \bibitem{Sasaki 2017} T. Sasaki, K. Kondo, Y. Akahama, S. Nakano, and T. Taniguchi, Jpn. J. Appl. Phys. \textbf{56}, 05FB06 (20170).
  
 
  
  

\bibitem[{\citenamefont{Giannozzi et~al.}(2009)\citenamefont{Giannozzi, Baroni,
  Bonini, Calandra, Car, Cavazzoni, Ceresoli, Chiarotti, Cococcioni, Dabo
  et~al.}}]{qe}
\bibinfo{author}{\bibfnamefont{P.}~\bibnamefont{Giannozzi}},
  \bibinfo{author}{\bibfnamefont{S.}~\bibnamefont{Baroni}},
  \bibinfo{author}{\bibfnamefont{N.}~\bibnamefont{Bonini}},
  \bibinfo{author}{\bibfnamefont{M.}~\bibnamefont{Calandra}},
  \bibinfo{author}{\bibfnamefont{R.}~\bibnamefont{Car}},
  \bibinfo{author}{\bibfnamefont{C.}~\bibnamefont{Cavazzoni}},
  \bibinfo{author}{\bibfnamefont{D.}~\bibnamefont{Ceresoli}},
  \bibinfo{author}{\bibfnamefont{G.~L.} \bibnamefont{Chiarotti}},
  \bibinfo{author}{\bibfnamefont{M.}~\bibnamefont{Cococcioni}},
  \bibinfo{author}{\bibfnamefont{I.}~\bibnamefont{Dabo}}, \bibnamefont{et~al.},
  \bibinfo{journal}{J. Phys. Condens. Matter} \textbf{\bibinfo{volume}{21}},
  \bibinfo{pages}{395502} (\bibinfo{year}{2009}).

\bibitem[{\citenamefont{Goedecker et~al.}(1996)\citenamefont{Goedecker, Teter,
  and Hutter}}]{SG}
\bibinfo{author}{\bibfnamefont{S.}~\bibnamefont{Goedecker}},
  \bibinfo{author}{\bibfnamefont{M.}~\bibnamefont{Teter}}, \bibnamefont{and}
  \bibinfo{author}{\bibfnamefont{J.}~\bibnamefont{Hutter}},
  \bibinfo{journal}{Phys. Rev. B} \textbf{\bibinfo{volume}{54}},
  \bibinfo{pages}{1703} (\bibinfo{year}{1996}).

\bibitem[{\citenamefont{Hartwigsen et~al.}(1998)\citenamefont{Hartwigsen,
  Goedecker, and Hutter}}]{CH}
\bibinfo{author}{\bibfnamefont{C.}~\bibnamefont{Hartwigsen}},
  \bibinfo{author}{\bibfnamefont{S.}~\bibnamefont{Goedecker}},
  \bibnamefont{and} \bibinfo{author}{\bibfnamefont{J.}~\bibnamefont{Hutter}},
  \bibinfo{journal}{Phys. Rev. B} \textbf{\bibinfo{volume}{58}},
  \bibinfo{pages}{3641} (\bibinfo{year}{1998}).

\bibitem[{\citenamefont{Perdew et~al.}(1996)\citenamefont{Perdew, Burke, and
  Ernzerhof}}]{PBE}
\bibinfo{author}{\bibfnamefont{J.~P.} \bibnamefont{Perdew}},
  \bibinfo{author}{\bibfnamefont{K.}~\bibnamefont{Burke}}, \bibnamefont{and}
  \bibinfo{author}{\bibfnamefont{M.}~\bibnamefont{Ernzerhof}},
  \bibinfo{journal}{Phys. Rev. Lett.} \textbf{\bibinfo{volume}{77}},
  \bibinfo{pages}{3865} (\bibinfo{year}{1996}).

\bibitem[{\citenamefont{Grimme}(2004)}]{Grimme}
\bibinfo{author}{\bibfnamefont{S.}~\bibnamefont{Grimme}}, \bibinfo{journal}{J.
  Comput. Chem.} \textbf{\bibinfo{volume}{25}}, \bibinfo{pages}{1463}
  (\bibinfo{year}{2004}).

\bibitem[{\citenamefont{Heyd et~al.}(2003)\citenamefont{Heyd, Scuseria, and
  Ernzerhof}}]{HSE}
\bibinfo{author}{\bibfnamefont{J.}~\bibnamefont{Heyd}},
  \bibinfo{author}{\bibfnamefont{G.~E.} \bibnamefont{Scuseria}},
  \bibnamefont{and}
  \bibinfo{author}{\bibfnamefont{M.}~\bibnamefont{Ernzerhof}},
  \bibinfo{journal}{J. Chem. Phys.} \textbf{\bibinfo{volume}{118}},
  \bibinfo{pages}{8207} (\bibinfo{year}{2003}).

\bibitem[{\citenamefont{Soluyanov and Vanderbilt}(2011)}]{soluyanov}
\bibinfo{author}{\bibfnamefont{A.~A.} \bibnamefont{Soluyanov}}
  \bibnamefont{and}
  \bibinfo{author}{\bibfnamefont{D.}~\bibnamefont{Vanderbilt}},
  \bibinfo{journal}{Phys. Rev. B} \textbf{\bibinfo{volume}{83}},
  \bibinfo{pages}{235401} (\bibinfo{year}{2011}).

\bibitem[{\citenamefont{Bhattacharjee and Waghmare}(2005)}]{joydeep}
\bibinfo{author}{\bibfnamefont{J.}~\bibnamefont{Bhattacharjee}}
  \bibnamefont{and} \bibinfo{author}{\bibfnamefont{U.~V.}
  \bibnamefont{Waghmare}}, \bibinfo{journal}{Phys. Rev. B}
  \textbf{\bibinfo{volume}{71}}, \bibinfo{pages}{045106}
  (\bibinfo{year}{2005}).

\bibitem[{\citenamefont{Fu and Kane}(2006)}]{fu}
\bibinfo{author}{\bibfnamefont{L.}~\bibnamefont{Fu}} \bibnamefont{and}
  \bibinfo{author}{\bibfnamefont{C.~L.} \bibnamefont{Kane}},
  \bibinfo{journal}{Phys. Rev. B} \textbf{\bibinfo{volume}{74}},
  \bibinfo{pages}{195312} (\bibinfo{year}{2006}).

\bibitem[{\citenamefont{Gresch et~al.}(2016)\citenamefont{Gresch, G.~Aut\'{e}s,
  M, Vanderbilt, Bernevig, and Soluyanov}}]{gresch}
\bibinfo{author}{\bibfnamefont{D.}~\bibnamefont{Gresch}},
  \bibinfo{author}{\bibfnamefont{O.~Y.} \bibnamefont{G.~Aut\'{e}s}},
  \bibinfo{author}{\bibfnamefont{.~T.} \bibnamefont{M}},
  \bibinfo{author}{\bibfnamefont{D.}~\bibnamefont{Vanderbilt}},
  \bibinfo{author}{\bibfnamefont{B.~A.} \bibnamefont{Bernevig}},
  \bibnamefont{and} \bibinfo{author}{\bibfnamefont{A.~A.}
  \bibnamefont{Soluyanov}}, \bibinfo{journal}{arXiv:1610.08983v1}
  (\bibinfo{year}{2016}).





\bibitem[{\citenamefont{Fei et~al.}(2015)\citenamefont{Fei, Tran, and
  Yang}}]{fei}
\bibinfo{author}{\bibfnamefont{R.}~\bibnamefont{Fei}},
  \bibinfo{author}{\bibfnamefont{V.}~\bibnamefont{Tran}}, \bibnamefont{and}
  \bibinfo{author}{\bibfnamefont{L.}~\bibnamefont{Yang}},
  \bibinfo{journal}{Phys. Rev. B} \textbf{\bibinfo{volume}{91}},
  \bibinfo{pages}{195319} (\bibinfo{year}{2015}).
  
  
  \bibitem{Akai 1989} T. Akai, S. Endo, Y. Akahama, K. Koto and Y. Maruyama, High Press. Res. \textbf{1}, 115 (1989).
  \bibitem{Svane 2012} S. Appalakondaiah, G. Vaitheeswaran, S. Leb\`egue, N. E. Christensen, and A. Svane, Phys. Rev. B \textbf{86}, 035105, (2012).
  
 \bibitem{Pisana} Simone Pisana, Michele Lazzeri, Cinzia Casiraghi, Kostya S. Novoselov, A. K. Geim, Andrea C. Ferrari and 
 Francesco Mauri, Nat. Mater., \textbf{6}, 198, (2007).
 
 
 \bibitem{Lazzeri} Michele Lazzeri, S. Piscanec, Francesco Mauri, A. C. Ferrari, and J. Robertson, Phys. Rev. B, \textbf{73}, 155426, (2006).
  
  
  
  

\bibitem[{\citenamefont{Mostofi et~al.}(2014)\citenamefont{Mostofi, Yates,
  Pizzi, Lee, Souza, Vanderbilt, and Marzari}}]{Wannier}
\bibinfo{author}{\bibfnamefont{A.~A.} \bibnamefont{Mostofi}},
  \bibinfo{author}{\bibfnamefont{J.~R.} \bibnamefont{Yates}},
  \bibinfo{author}{\bibfnamefont{G.}~\bibnamefont{Pizzi}},
  \bibinfo{author}{\bibfnamefont{Y.-S.} \bibnamefont{Lee}},
  \bibinfo{author}{\bibfnamefont{I.}~\bibnamefont{Souza}},
  \bibinfo{author}{\bibfnamefont{D.}~\bibnamefont{Vanderbilt}},
  \bibnamefont{and} \bibinfo{author}{\bibfnamefont{N.}~\bibnamefont{Marzari}},
  \bibinfo{journal}{Comput. Phys. Commun.} \textbf{\bibinfo{volume}{185}},
  \bibinfo{pages}{2309 } (\bibinfo{year}{2014}).

\end{thebibliography}

\end{document}